\documentclass[12pt]{article}
\usepackage{amssymb}
\usepackage[dvips]{graphicx}
\usepackage{amsmath}
\usepackage{amsfonts, amscd}

\usepackage{amssymb}
\usepackage{latexsym}

\newtheorem{definition}{\sc Definition}[section]
\newtheorem{example}{\sc Example}[section]

\newcommand{\BF}[1]{{\bf#1}}

\newcommand{\RM}[1]{{\rm#1}}

\renewcommand{\SS}{\subseteq}

\newcommand{\NW}[1]{{\em#1}}
\newcommand{\UP}[1]{{\rm#1}}

\newcommand{\noqed}{\renewcommand{\qed}{}} 

\newcommand{\qed}{~~~\mbox{$\Box$}}                 
\newcommand{\prfn}{Proof}                 
\newenvironment{prf}{{\it \prfn.}}{\qed\smallbreak} 

\def\bprf{\if!=} 
\let\eprf\fi

\newenvironment{trm}[1]{\smallbreak{\bf Theorem #1.}\it}{\smallbreak}
\newenvironment{lem}[1]{\smallbreak{\bf Lemma #1.}\it}{\smallbreak}
\newenvironment{prp}[1]{\smallbreak{\bf Proposition #1.}\it}{\smallbreak}

\newcommand{\rf}[1]{{\rm(#1)}}  
\newcommand{\ct}[1]{\cite{#1}}          

\newcommand{\Iff}{\;\iff\;}
\newcommand{\iffdef}{\buildrel{\RM{def}}\over\iff}
\newcommand{\Iffdef}{\;\iffdef\;}
\newcommand{\imp}{\;\Longrightarrow\;}
\newcommand{\Imp}{\;\imp\;}
\newcommand{\eqdef}{\buildrel{\RM{def}}\over=}

\newcommand{\cls}[1]{{\BF{#1}}}     

\newcommand{\MVT}{\cls{MVT}}                       
\newcommand{\Set}{\cls{Set}}                       

\newcommand{\UU}{\BF{U}}
\newcommand{\CC}{{\bf{C}}}
\newcommand{\DD}{{\bf{D}}}

\newcommand{\EC}[1]{{\bf#1}}               

\newcommand {\spc}[1]{{\cal #1}}

\newcommand{\A}{\spc{A}}
\newcommand{\B}{\spc{B}}
\newcommand{\C}{\spc{C}}
\newcommand{\R}{\spc{R}}    
\newcommand{\D}{\spc{D}}    
\newcommand{\U}{\spc{U}}    

\newcommand{\LI}{\preccurlyeq}                
\newcommand{\BI}{\succcurlyeq}                
\newcommand{\EI}{\sim}                        

\newcommand{\BS}{\sqsupseteq}                 

\newcommand{\BQ}{\gg}                         
\newcommand{\EQ}{\approx}                     

\newcommand{\GE}{\geqslant}  
\newcommand{\LE}{\leqslant}  

\newcommand{\HP}{\vartriangleright}          

\newcommand{\cUp}{{\textstyle\bigcup}}

\newcommand{\CE}[1]{\bar{#1}}

\newcommand{\Ob}{\mathop{\RM{Ob}}\nolimits}
\newcommand{\Ar}{\mathop{\RM{Ar}}\nolimits}
\newcommand{\Opt}{\mathop{\RM{Opt}}\nolimits}

\newcommand{\smalltimes}{\mathbin{\mathchoice%
           {\raise.2ex\hbox{$\scriptstyle\times$}}%
           {\raise.2ex\hbox{$\scriptstyle\times$}}%
           {\raise.0ex\hbox{$\scriptscriptstyle\times$}}%
           {\raise.0ex\hbox{$\scriptscriptstyle\times$}}}}

\renewcommand{\(}{\left(\big.}
\renewcommand{\)}{\right)}

\renewcommand{\.}{\circ}                                   
\renewcommand{\*}{\smalltimes}                             

\renewcommand{\:}{\colon}                                  
\renewcommand{\>}{\to}                                     

\renewcommand{\#}{\times}                                  

\newcommand{\q}{\hskip.7em\relax}

\newcommand{\<}[1]{_{{}_{\!\scriptstyle#1\!}}}          

\newcommand{\It}[1]{\mbox{\UP{(#1)}}}

\renewcommand{\=}{\\&=&}

\newcommand{\EMPTYSET}{\varnothing}

\newcommand{\IF}{&\mbox{if}&}
\def\CASE#1&#2;#3&#4;{ \left\{\begin{array}{lll}
                         #1,\IF#2,\\[2mm]#3,\IF#4
                       \end{array}\right. }

\makeatletter

\begingroup
  \catcode`\|=13
  \gdef\SET{\catcode`\|=13
\def|{\setbox0=\hbox\bgroup$\displaystyle\;\left\vert\;\mathstrut}\SET@}

\gdef\SET@#1{{\left\{               
  \setbox1=\hbox{$\let|=\;#1$}
  #1 \vphantom{{\let|=\;\displaystyle#1}}
  \right.$\egroup\ht0=1.2\ht0\box0
  \right\} }}
\endgroup

\makeatother

\def\TPL#1{\left\langle#1\right\rangle}

\let\UL=\_

\def\_{\nobreak\hskip0pt-\allowbreak\hskip0pt} 

\newcommand{\Meas}{\cls{Meas}}

\newcommand{\Z}{{\spc{Z}}}  
\renewcommand{\P}{{\spc{P}}}
\newcommand{\Q}{{\spc{Q}}}

\newcommand{\al}{\alpha}
\newcommand{\la}{\lambda}
\newcommand{\rh}{\rho}
\newcommand{\si}{\sigma}
\newcommand{\de}{\delta}
\newcommand{\et}{\eta}
\newcommand{\ga}{\gamma}

\def\Put(#1)#2{\put(#1){\raisebox{-.8ex}{\makebox[0pt][c]{$#2$}}}}
\def\Vec(#1;#2;#3){\put(#1){\vector(#2){#3}}}

\def\Dia(#1,#2)#3{%
\countdef\y=1\y=#2%
\countdef\dy=2\dy=2
\advance\y by\dy\advance\y by\dy%
\begin{picture}(#1,\number\y)(0,-\number\dy)%
#3%
\end{picture}%
}

\newcommand{\Hom}{{\rm Hom}}
\newcommand{\id}{{\rm id}}
\newcommand{\Sets}{{\rm \bf Sets}}
\newcommand{\Top}{{\rm \bf Top}}
\newcommand{\Cat}{{\rm \bf Cat}}
\newcommand{\X}{{\bf X}}
\newcommand{\Y}{{\bf Y}}

\begin{document}

\setcounter{page}{1}

\begin{center}

{\Large \bf 
 Method of Additional Structures on the Objects of a Monoidal Kleisli Category as a Background for Information Transformers Theory}

\bigskip

Peter V. Golubtsov


Moscow State Lomonosov University


Department of Physics, Moscow State University \\

119899, Moscow, Russia \\

E-mail P\UL V\UL G@mail.ru

\bigskip
S.S. Moskaliuk

\medskip

Bogolyubov Institute for Theoretical Physics\\
Metrolohichna Str., 14-b, Kyiv-143, Ukraine, UA-03143\\
e-mail: mss@bitp.kiev.ua

\end{center}

\bigskip

\begin{center}
{\bf Abstract}
\end{center}

\medskip

Category theory provides a compact method of encoding 
mathematical structures in a uniform way, thereby enabling the 
use of general theorems on, for example, equivalence and 
universal constructions.
In this article
 we develop the method of additional structures on the objects of a monoidal Kleisli category. It is proposed to consider any uniform class of
information transformers (ITs) as a family of morphisms of a
category that satisfy certain set of axioms. This makes it
possible to study in a uniform way different types of ITs, e.g.,
statistical, multivalued, and fuzzy ITs. Proposed axioms define a
category of ITs as a monoidal category that contains a subcategory
(of deterministic ITs) with finite products. Besides, it is shown
that many categories of ITs can be constructed as Kleisli
categories with additional structures.

\newpage
\section{Introduction}

Currently the growing interest is attracted to various
mathematical ways of describing uncertainty, most of them being
different from the probabilistic one, (e.g., based on the
apparatus of fuzzy sets). For adequate theoretical study of the
corresponding ``nonstochastic'' systems of information
transforming and, in particular, for the study of important
notions, such as sufficiency, informativeness, etc., we need to
develop an approach general enough to describe different classes
of information transforming systems in a uniform way.

It is convenient to consider different systems that take place in
information acquiring and processing  as particular cases of so-called
\NW{information transformers} (ITs). Besides, it is useful to work with families of
ITs in which certain operations, e.g., \NW{sequential} and \NW{parallel
compositions} are defined.

It was noticed fairly long ago~%
[1--5],
that the adequate algebraic
structure for describing information transformers (initially for
the study of statistical experiments) is
the structure of \NW{category}~%
[6--9].

\begin{definition}
 A category is a quadruple $(\Ob, \Hom, \id, \circ)$
consisting of:

(Cl) a class $\Ob$ of objects;

(C2) for each ordered pair $(A, B)$ of objects a set $\Hom(A, B)$
of morphisms;

(C3) for each object $A$ a morphism $\id_A \in \Hom(A,A)$,
the identity of $A$;

(C4) a composition law associating to each pair of morphisms
$f\in  \Hom(A, B)$ and  $g \in \Hom(B, C)$ a morphism
$g\circ f \in  \Hom(A, C)$;

\noindent
which is such that:

(Ml) $h\circ (g\circ f) = (h\circ g)\circ f$ for all
$f \in \Hom(A,B)$, $g \in \Hom(B,C)$ and $h \in \Hom(C,D)$;

(M2) $\id_B \circ f =f\circ \id_A =f$  for all $f \in  Hom(A, B)$;

(M3) the sets $\Hom(A, B)$ are pairwise disjoint.
\end{definition}

This last axiom is necessary so that given a morphism we can identify
its domain $A$ and codomain $B$, however it can always be satisfied
by replacing $\Hom(A,B)$ by the set $\Hom(A,B) \times (\{A\},\{B\})$.

A morphism $a\:\A\>\B$ is called \NW{isomorphism} if there exists a morphism $b\:\B\>\A$
such that $a\.b=i\<\B$ and $b\.a=i\<\A$. In this case objects $\A$ and $\B$ are called
\NW{isomorphic}.

Morphisms $a\:\D\>\A$ and $b\:\D\>\B$ are called \NW{isomorphic} if there exists an
isomorphism $c\:\A\>\B$ such that $c\.a=b$.

An object $\Z$ is called \NW{terminal} object if for any object $\A$ there exists a unique morphism
from $\A$ to $\Z$, which is denoted $z\<{\A}\:\A\>\Z$ in what follows.

A category $\DD$ is called a \NW{subcategory} of a category $\CC$ if $\Ob(\DD)\SS\Ob(\CC)$,
$\Ar(\DD)\SS\Ar(\CC)$, and morphism composition in $\DD$ coincide with their composition in
$\CC$.

It is said that a category has (pairwise) products if for every pair of objects $\A$ and $\B$ there
exists their \NW{product}, that is, an object $\A\#\B$ and a pair of morphisms
$\pi\<{\A,\B}\:\A\#\B\>\A$ and $\nu\<{\A,\B}\:\A\#\B\>\B$, called projections, such that for any
object $\D$ and for any pair of morphisms $a\:\D\>\A$ and $b\:\D\>\B$ there exists a unique
morphism $c\:\D\>\A\#\B$, satisfying the following conditions:
\begin{equation} 
  \pi\<{\A,\B}\.c = a, \qquad\nu\<{\A,\B}\.c = b.
\end{equation}

We call such morphism $c$ the \NW{product of morphisms} $a$ and $b$ and denote it $a*b$.

It is easily seen that existence of products in a category implies the following equality:
\begin{equation} 
  (a*b)\.d = (a\.d)*(b\.d).
\end{equation}

In a category with products, for two arbitrary morphisms $a\:\A\>\C$ and $b\:\B\>\D$ one can
define the morphism $a\*b$:
\begin{equation}  
  a\*b\:\A\#\B\>\C\#\D, \qquad
a\*b\eqdef (a\.\pi\<{\A,\B})*(b\.\nu\<{\A,\B}).
\end{equation}

This definition and{}~\rf{1} obviously imply that the morphism $c=a\*b$ satisfy the following
conditions:
\begin{equation} 
  \pi\<{\C,\D}\.c = a\.\pi\<{\A,\B}, \qquad
\nu\<{\C,\D}\.c = b\.\nu\<{\A,\B}.
\end{equation}
Moreover, $c=a\*b$ is the only morphism satisfying conditions{}~\rf{4}.

It is also easily seen that{}~\rf{2} and{}~\rf{3} imply the following equality:
\begin{equation} 
  (a\*b)\.(c*d) = (a\.c)*(b\.d).
\end{equation}

Suppose $\A\#\B$ and $\B\#\A$ are two products of objects $\A$ and $\B$ taken in different order. By the properties of products, the objects $\A\#\B$ and $\B\#\A$ are isomorphic and the
natural isomorphism is
\begin{equation} 
  \si\<{\A,\B}\:\A\#\B\>\B\#\A, \qquad
  \si\<{\A,\B}\eqdef \nu\<{\A,\B}*\pi\<{\A,\B}.
\end{equation}

Moreover, for any object $\D$ and for any morphisms $a\:\D\>\A$ and $b\:\D\>\B$, the
morphisms $a*b$ and $b*a$ are isomorphic, that is,
\begin{equation} 
  \si\<{\A,\B}\.(a*b) = b*a.
\end{equation}

Similarly, by the properties of products, the objects $(\A\#\B)\#\C$ and $\A\#(\B\#\C)$ are
isomorphic. Let
\[
  \al\<{\A,\B,\C}\: (\A\#\B)\#\C\> \A\#(\B\#\C)
\]
be the corresponding natural isomorphism. Its ``explicit'' form is:
\begin{equation}  
  \al\<{\A,\B,\C} \eqdef
  (\pi\<{\A,\B}\.\pi\<{\A\#\B,\C})*\((\nu\<{\A,\B}\.\pi\<{\A\#\B,\C})*\nu\<{\A\#\B,\C}\).
\end{equation}

Then for any object $\D$ and for any morphisms $a\:\D\>\A$, $b\:\D\>\B$, and $c\:\D\>\C$ we
have
\begin{equation} 
  \al\<{\A,\B,\C}\.\((a*b)*c\) = a*(b*c).
\end{equation}

{\sc Examples.}

{\bf 1.1.}  The classic example is $\Sets$, the category with sets as objects
and functions as morphisms, and the usual composition of functions as
composition.
But lots of the time {\it in mathematics} one is some category or other,
e.g.:

{\bf Vect$_k$} ---  vector spaces over a field $k$ as objects;
  $k$-linear maps as morphisms;

{\bf Group} --- groups as objects, homomorphisms as morphisms;

$\Top$ --- topological spaces as objects, continuous functions
  as morphisms;

{\bf Diff} --- smooth manifolds as objects, smooth maps as 
morphisms;

{\bf Ring} ---  rings as objects, ring homomorphisms as 
morphisms;


\noindent
or {\it in physics}:

{\bf Symp} --- symplectic manifolds as objects, 
symplectomorphisms as morphisms;

{\bf Poiss} ---  Poisson manifolds as objects, Poisson maps as 
morphisms;

{\bf Hilb} ---  Hilbert spaces as objects, unitary operators as 
morphisms.

{\bf 1.2.} The typical way to think about symmetry is with the 
concept of a "group". But  to get a concept of symmetry that's 
really up to the demands put on it by modern mathematics and 
physics, we need --- at the very least ---  to work with a 
"category" of symmetries, rather than a group of symmetries.

To see this, first ask: what is a category with one object? It is 
a --- {\it "monoid"}.  The "usual" definition of a monoid is like
this: a set $M$ with an associative binary product and a unit 
element 1 such that $al = la = a$ for all $a$ in $M$. Monoids 
abound in mathematics; they are in a sense the most primitive 
interesting algebraic structures.

To check that a category with one object is "essentially just a monoid",
note that if our category $C$ has one object $x$, the set
$\Hom(x,x)$  of all morphisms from $x$ to $x$ is indeed a set with an
associative binary product, namely composition, and a unit element,
namely $\id_x$.

How about categories in which every morphism is invertible?
We  say a morphism $f: x \to  y$ in a category has inverse $g: y \to x$
if $f\circ g = \id_y$ and $g\circ f = \id_x$. Well, a category
in which every morphism is invertible is
called a {\it "groupoid"}.

Finally, a  group is a category with one object in which every morphism
is invertible. It's both a monoid and a groupoid!

When we use groups in physics to describe symmetry, we think of each
element $g$ of the group $G$ as a "process". The element $1$ corresponds
to the "process of doing nothing at all". We can compose processes $g$
and $h$ --- do $h$ and then $g$ --- and get the product $g\circ h$.
Crucially, every process $g$ can be "undone" using its inverse 
$g^{-1}$.

So: a monoid is like a group, but the "symmetries" no longer need be
invertible; a category is like a monoid, but the "symmetries" no longer
need to be composable.

{\bf 1.3.}  The operation of "evolving initial data from one
spacelike slice to another" is a good example of a "partially defined"
process: it only applies to initial data on that particular spacelike
slice. So dynamics in special or general relativity is most naturally
described {\it using groupoids}. Only after pretending that all the
spacelike slices are the same can we pretend we are using a group.
It is very common to pretend that groupoids are groups, since groups
are more familiar, but  often insight is lost in the process.
Also, one can only pretend a groupoid is a group if all its objects
are isomorphic. Groupoids really are more general.

In the work [10] we undertake an attempt to formulate the method of categorical extension of the theory of a group $G$ as follows:

Let $G$ be a group. Then $G$ is merely the visible part of a certain category $K$ which is invisible to the naked eye. More precisely, there exists a certain category $K$ (the {\it train} of the group $G$) such that the group itself is the automorphism group of a certain object $V$, while the semigroup $\Gamma$ is the semigroup of endomorphisms of this same object. Furthermore, each representation $\rho$ of $G'$ on a space $H$ can be extended to a representation of the category $K$. In other words, for each objects $W$ of the category $K$ we can construct a linear space $T(W)$ and for each
 morphism $P: W \to W'$ we can construct a linear operator $\tau(P): T(W) \to T(W')$ such that for any morphisms $P: W \to W'$ and $Q: W' \to W''$ we have 

$$
 \tau(QP) = \tau(Q)\tau(P)
$$
 
with $T(V) = H$, and for all $g \in G$ the operators $\tau(g)$ and $\rho(g)$ are the same.

We note that all the spaces $T(W)$ and all the operators $\tau(p)$ ``grow out of'' the one and only representation $\rho$ of $G$ and the one and only space $H$.
 
So: in contrast to a set, which consists of a static collection of
"things", a category consists not only of objects or "things" but also
morphisms which can viewed as "processes" transforming one thing into
another. Similarly, in a 2-category, the 2-morphisms can be regarded as
"processes between processes", and so on. The eventual goal of basing
mathematics upon omega-categories is thus to allow us the freedom to
think of any process as the sort of thing higher-level processes can go
between. By the way, it should also be very interesting to consider
"$\mathbb{Z}$-categories" (where $\mathbb{Z}$ denotes the integers),
having $j$-morphisms not only for $j = 0,1,2,...$ but also for negative
$j$. Then we may also think of any thing as a kind of process.

\begin{definition}
 Let ${\bf X}$ and ${\bf Y}$ be two categories.
 A functor from ${\bf X}$ to ${\bf Y}$  is a family of functions $F$
 which associates to each object $A$ in ${\bf X}$ an object
 $FA$ in ${\bf Y}$  and to each morphism $f \in \Hom_{\bf X}(A, B)$ a
 morphism $Ff \in \Hom_{\bf Y}(FA, FB)$, and which is such that:

(FI) $F(g \circ f) = Fg \circ  Ff$ for all $f \in \Hom_{\bf X}(A, B)$
and $g\in \Hom_{\bf Y}(B, C)$;

(F2) $F\, \id_A = \id_{FA}$ for all $A \in \Ob ({\bf X})$.
\end{definition}

There is the definition of left and right adjoint functors.
In the following we shall need two such adjoint constructions.
First, in a given category the left adjoint of the diagonal functor
(if it exists) is called the coproduct and the right
adjoint (if it exists) is called the product: in $\Sets$ the product is
the Cartesian product and the coproduct is the disjoint union.
Second, let the category ${\bf X}$ be concrete over some
category ${\bf A}$ in the sense that there exists a faithful functor
$U$ from ${\bf X}$ to ${\bf A}$, usually called the forgetful functor.
The left adjoint to this functor (if it exists) is then called the free
functor. A standard example is the forgetful functor from complete metric
spaces to metric spaces, whose left adjoint in the completion functor.
On the next higher level of abstraction the notion of a {\it 
natural} transformation is settled. It is a kind of a function 
between functors and is defined as follows.

\begin{definition}
Let $F:\X\rightarrow \Y$ and $G:\X\rightarrow \Y$ be two 
functors. A {\it natural transformation} $\alpha: F\rightarrow G$ 
is given by the following data.

For every object $A$ in $\X$ there is a morphism $\alpha_A: F(A) 
\rightarrow G(A)$ in $\Y$ such that for every morphism $f: 
A\rightarrow  B$ in $\X$ the following diagram is commutative
\begin{equation*}
\begin{array}{ccc}
{F(A)} & \overset{\alpha_A}{\longrightarrow } & {G(A)}\\
{F(f) \downarrow } &  & {{\;}{\;\downarrow G(f)}} \\
{F(B)} & \overset{\alpha_B}{\longrightarrow } & {G(B)}.\\
\end{array}
\end{equation*}

\end{definition}

Commutativity means (in terms of equations) that the following 
compositions of morphisms are equal: $G(f)\circ 
\alpha_A=\alpha_B\circ F(f)$. 

The morphisms $\alpha_A$, $A\in {\rm Obj}(A)$, are called the 
components of the natural transformation $\alpha$.

{\sc Examples.}

{\bf 1.4.} So, we can certainly speak, as before, of the 
"equality" of categories. We can also speak of the "isomorphism" 
 of categories: an isomorphism between ${\bf C}$ and ${\bf D}$ is 
a functor $F: {\bf C} \to {\bf D}$ for which there is an inverse 
functor $G: {\bf  D}\to {\bf  C}$.  I.e., $FG$ is the identity 
functor on ${\bf  C}$ and $GF$ the identity on ${\bf D}$, where 
we define the composition of functors in the obvious way. But 
because we also have natural transformations, we can also define 
a subtler notion, the "equivalence" of categories.  An 
equivalence is a functor $F: {\bf  C} \to {\bf  D}$ together with 
a functor $G: {\bf  D}\to  {\bf C}$ and natural isomorphisms $a: 
FG \to 1_C$ and $b: GF \to 1_D$.  A "natural isomorphism" is a 
natural transformation which has an inverse.

{\bf 1.5.} As we can "relax" the notion of equality to the notion of
isomorphism when we pass from sets to categories, we can relax the
condition that $FG$ and $GF$ equal identity functors to the condition
that they be isomorphic to identity functor when we pass from categories
to the 2-category $\Cat$. We need to  have the natural transformations
to be able to speak of functors being isomorphic, just as we needed
functions to be able to speak of sets  being isomorphic.
In fact, with each extra level in the theory of $n$-categories, we will
be able to come up with a still more refined notion of "$n$-equivalence"
in this way.

Analysis of general
properties for the classes of linear, multivalued, and fuzzy
information transformers, studied in [5, 11--18],
allowed to extract general features shared by all these classes.
Namely, each of these classes can be considered as a family of
morphisms in an appropriate category, where the composition of
information transformers corresponds to their ``consecutive
application.'' Each category of ITs (or IT-category) contains a
subcategory (of so called, \NW{deterministic} ITs) that has products.
Moreover, the operation of morphism product is extended in a
``coherent way'' to the whole category of ITs.

The works [19--22]
undertook an attempt to
formulate the method of additional structures as a set of ``elementary'' axioms for a category,
which would be sufficient for an abstract expression of the basic
concepts of the theory of information transformers and for study
of informativeness, decision problems, etc. This paper proposes
another, significantly more compact axiomatic for a category of
ITs. According to  the method of additional structures on the objects of  a category of ITs it is defined in
effect as a \NW{monoidal} category
[6, 8],
containing a subcategory (of
\NW{deterministic} ITs) with finite products.

Among the basic concepts connected to information transformers there
is one that plays an important role in the uniform construction of a wide
spectrum of IT-categories --- the concept of \NW{distribution}. Indeed, fairly often
an IT $a\:\A\>\B$ can be represented by a mapping from $\A$ to the ``space
of distributions'' on $\B$ (see, e.g.,
[11--18]). For example, a probabilistic
transition distribution (an IT in the category of stochastic ITs) can be
represented by a certain measurable mapping from $\A$ to the space of
distributions on $\B$. This observation suggests to construct a category of ITs
as a \NW{Kleisli} category 
 [6,23],
 arising from the following components: an
obvious category of deterministic ITs; a functor that takes an object $\A$ to
the object of ``distributions'' on $\A$; and a natural transformation of
functors, describing an ``independent product of distributions''.

It appears that rather general axiomatic theory, obtained this
way, makes it possible to express in terms of IT-categories basic
concepts for information transformers and to derive their main
properties.

Of cause, the most developed theory of uncertainty is probability
theory (and statistics, based on probability). Certainly,
mathematical statistics accumulated a rich conceptual experience.
It introduced and deeply investigated such notions as joint and
conditional distributions, independence, sufficiency, and others.

At the same time, it appears that all these concepts have very
abstract meaning and hence, they can be treated in terms of
alternative (i.e., not probabilistic) approaches to the
description of uncertainty. In fact, the basic notions of
probability theory and statistics, as well as the methodology and
results,
are easily extended to other theories dealing with uncertainty. In
[11--18]
 it is shown that a rather substantive decision theory may be constructed even on the very moderate
basis of multivalued or fuzzy maps.

The approach developed in this paper allows to express easily in
terms of IT-categories such concepts as distribution, joint and
conditional distributions, independence, and others. It is shown
that on the basis of these concepts it is possible to formulate
fairly general statement of decision-making problem with a prior
information, which generalizes the Bayesian approach in the theory
of statistical decisions. Moreover, the \NW{Bayesian} principle,
derived below, like its statistical
prototype [24],
reduces the problem of optimal decision strategy
construction to a significantly simpler problem of finding optimal
decision for a posterior distribution.

Among the most important concepts in categories of ITs  is the
concept of (relative) \NW{informativeness} of information transformers.
There are two different approaches to the concept of
informativeness.

One of these approaches is based on analyzing the ``relative
positions'' of information transformers in the corresponding
mathematical structure. Roughly speaking, one information
transformer is regarded as more informative than another one if
with the aid of an additional information transformer the former
one can be  ``transformed'' to an IT, which is similar to (or more
``accurate'' than) the latter one. In fact, this means that all
the information that can be obtained from the latter information
transformer can be extracted from the former one as well.

The other approach to informativeness is based on treating
information transformers as data sources for decision-making
problems. Here, one information transformer is said to be
semantically more informative than another if it provides better
quality of decision making. Obviously, the notion of semantical
informativeness depends on the class of decision-making problems
under consideration.

In the classical researches of Blackwell [25, 26]
the correspondence between
informativeness (Blackwell sufficiency) and semantical
informativeness (Blackwell informativeness) were investigated in a
statistical context. These studies were extended
by Morse, Sacsteder, and Chentsov [1--4]
who applied the category theory techniques to their studies of statistical systems.

It is interesting, that under very general conditions the
relations of informativeness and semantical informativeness (with
respect to a certain class of decision-making problems) coincide.
Moreover, in some categories of ITs it is possible to point out
{\em one} special decision problem, such that the resulting
semantical informativeness coincides with informativeness.

Analysis of classes of equivalent (with respect to
informativeness) information transformers shows that they form a
partially ordered Abelian monoid with the smallest (also neutral)
and the largest elements.

One of the objectives of this paper is to show that the basic
constructions and propositions of probability theory and
statistics playing the fundamental role in decision-making
problems have meaningful counterparts in terms of IT-categories.
Furthermore, some definitions and propositions (for example, the
notion of conditional distribution and the Bayesian principle) in
terms of IT-categories often have more transparent meanings. This
provides an opportunity to look at the well known results from a
different angle. What is even more significant, it makes it
possible to apply the methodology of statistical decision-making
in an alternative (not probabilistic) context.

Approaches, proposed in this work may provide a background for
construction and study of new classes of ITs, in particular, dynamical
nondeterministic ITs, which may provide an adequate description for
information flows and information interactions evolving in time. Besides, a
uniform approach to problems of information transformations may be useful
for better understanding of information processes that take place in complex
artificial and natural systems.

\section{The method of additional structures
on the objects of a category}

\subsection{Basic definitions}

 To use the categorical language more effectively
we introduce general concept of an additional structure on
objects of a category. This is the concept  of concrete category
but over any category [19--22].

In a category, two objects $x$ and $y$ can be equal or not equal, but 
they can be {\it isomorphic} or not, and if they are isomorphic,
they can be isomorphic in many different ways.
An isomorphism between $x$ and $y$ is simply a morphism
$f: x \to  y$ which has an inverse $g: y \to  x$, such that
$f\circ g =\id_y$ and  $g\circ f =\id_x$.

In the category $\Sets$  an isomorphism is just a one-to-one and
onto function, i.e. a bijection. If we know two sets $x$ and $y$
are isomorphic we know that
they are "the same in a way", even if they are not equal. But
specifying an isomorphism $f: x \to y$ does more than say $x$ and
$y$ are the same in a way; it specifies a {\it particular way} to
regard $x$ and $y$ as the same.

In short, while equality is a yes-or-no matter, a mere {\it property},
an isomorphism is a {\it structure}. It is quite typical, as we climb the
categorical latter (here from elements of a set to objects of a
category) for properties to be reinterpreted as structures.

\begin{definition}
We tell that a functor $F:{\cal C}\to {\cal C}^{\prime }$ define a {\it 
additional ${\cal C}-$structure on objects of the category} ${\cal C}%
^{\prime }$ if

\begin{enumerate}
\item  $\forall X,Y\in Ob({\cal C})$ the map $F:{\cal C}(X,Y)\to 
{\cal C}^{\prime }(F(X),F(Y))$ is injective,

\item  $\forall X\in Ob({\cal C}),Y\in Ob({\cal C}^{\prime })$ and an
isomorphism $u:Y\to F(X)$ there is an object $\tilde Y\in Ob$ and an
isomorphism $\tilde u:\tilde Y\to X$ such that $F(\tilde Y)=Y$ and $F(\tilde
u)=u$.
\end{enumerate}

Such functor is called a {forgetful} functor.
\end{definition}

Almost all usual mathematical structures are structure on sets in this sense
and there are corresponding forgetful functors to the category
$\Sets$ of sets.

A forgetful functor $F:{\cal C}\to M({\cal C}^{\prime })$ defines a
${\cal C}$-structure {\it on morphisms of the category} ${\cal C^{\prime}}$.

For our general structures we can define usual construction:

\begin{enumerate}
\item[--]  inverse and direct images of structures;
\item[--]  restrictions on subobjects,
\item[--]  different products of structures.
\end{enumerate}

We can define the category $Str({\cal C})$ of forgetful functors to the
category ${\cal C}$. It is a full subcategory of the category $Cat/{\cal C}$
of all categories over ${\cal C}$.

Some properties of structures (= forgetful functors):

\begin{enumerate}
\item[--]  In the category $Str({\cal C})$ the (bundle) product 
always exists.  It gives a ``union'' structures.  

\item[--]  Any 
functor $f:{\cal C}\to {\cal C}^{\prime }$ transfers structures 
to inverse direction, i.e. it defines the functor 
\begin{equation*}
f^{*}:Str({\cal C}^{\prime })\to Str({\cal C}):F\mapsto f^{*}F.
\end{equation*}
\item[--]  For a forgetful functor $F:{\cal C}\to {\cal 
C}^{\prime }$ the functors 
\begin{equation*} 
(F\circ 
)=Funct(id,F):Funct({\cal B},{\cal C})\to Funct({\cal B}, {\cal 
C}^{\prime }) 
\end{equation*} 
\begin{equation*} 
(\circ 
F)=Funct(F,id):Funct({\cal C}^{\prime },{\cal B})\to Funct({\cal 
C},{\cal B}) 
\end{equation*}
are forgetful functors.  
\item[--]  One of 
constructions which transfers structure $F:{\cal 
C}\to Sets$ defined on sets to objects of any category ${\cal B}$, 
is the functor 
\begin{equation*} 
h:{\cal B}\to Funct({\cal 
B}^{\circ },Sets):B\mapsto h_B.  \end{equation*} 
Thus we have 
\begin{equation*}
\begin{array}{ccc}
{h_B^{*}{\cal C}} & \longrightarrow  & {\cal C} \\
{\downarrow } &  & {{\;}{\;\downarrow F}} \\
{{\cal B}^{\prime }} & \overset{h_B}{\longrightarrow } & {Sets}
\end{array}
\end{equation*}

\item[--]  If a functor $A:{\cal B}\to {\cal C}$ is injective on
morphisms (the condition (1) in the definition of forgetful 
functor) then a forgetful functor $F:{\cal B}^{\prime }\to {\cal 
C}$ and an equivalence $i:{\cal B}\to {\cal B}^{\prime }$  
exist, such that the following diagram is commutative 
\begin{equation*} 
\begin{array}{ccc} 
{\cal B} & \longrightarrow & {\cal C} \\ 
{\downarrow } & {\;\qquad \nearrow } _F &  \\ 
{{\cal B}^{\prime }} &  &
\end{array}
\end{equation*}
\end{enumerate}

\subsection{Structures on Topological Spaces}

Among of structures on topological spaces we can select that, which
is compatible with the topology. Let $Top$ be a category of some 
topological spaces with a forgetful functor $F:Top\to Sets.$

The categories associated with a topological space $T\in Ob(Top)$ 
as follows:

\begin{enumerate}
\item[--]  The category ${\cal T}(T),$ where $Ob({\cal T}(T))$ is 
the set of all open subsets of $T$, and $Mor({\cal T}(T))$ is all 
their inclusions.

\item[--]  The category (pseudogroup) ${\cal P}(T)$, where 
$Ob({\cal P}(T))$ is the set of all open subsets of $T$, and 
$Mor({\cal P}(T))$ is all their homeomorphisms.  \end{enumerate}

Functors ${\cal T}(T)^{\circ }\to Set$ are called {\it presheaves} of sets
on $T$. Some of them are called sheaves. Thus we have the 
inclusions 
\begin{equation*} 
Sh(T)\subset Presh(T)\subset 
Funct({\cal T}(T),Sets).  \end{equation*}

A {\it Grothendieck topology} on a category is defined by saying which
families of maps into an object constitute a {\it covering} of the
object and certain axioms are fulfill. A category together with a
Grothendieck topology on it is called a {\it site}. For a site ${\cal C}$
one define the full subcategory $Sh({\cal C})\subset Presh({\cal C})=Funct(%
{\cal C}^{\circ},Set).$ The objects of $Funct({\cal C}^{\circ},Set)$ are
called {\it presheaves} on the site ${\cal C},$ and the objects of $Sh({\cal %
C})$ are called {\it sheaves} on ${\cal C}.$

For any category there exists the finest topology such that the all
representable presheaves are sheaves. It is called the {canonical}
Grothendieck topology. {\it Topos} is a category which is equivalent to the
category of sheaves for the canonical topology on them. 

Hence, the topology is already transfered on a category so now 
it is natural to 
consider on language of toposes and sheaves all questions connected to local properties.

Here we shall not consider local structures  on
toposes in general, and we shall restrict ourselves with the 
consideration of the elementary case of the category $\Top$.

\begin{definition}
A structure defined by a forgetful functor $f:{\cal C}\to Top$ is called a
{\bf local structure} if

 $\forall C\in Obj({\cal C})$ and any inclusion map $i:U\to f(C)$ of
the open subset $U$ an object $\tilde U\in Ob({\cal C})$ and a morphism
$\tilde i\in {\cal C}(\tilde U,C)$ exist such that $f(\tilde U)=U$
$f(\tilde  i)=i.$
This ${\cal C}-$structure $\tilde U$ is denoted by $C|U$ and called a
{\bf restriction} of $C$ on $U$.
\end{definition}

In other words we can restrict ourselves with local structures on 
open subsets.

For a local structure $F:{\cal C}\to Top$ and each object $X\in Obj(Top)$
there is the presheaf of categories
\begin{equation*}
{\cal T}(X)^{\circ}\to Cat:U\mapsto F^{-1}(U,id_U).
\end{equation*}
Often this presheaf is a sheaf.

\subsection{Structures on Smooth Manifolds}

Let ${\cal M}$ be the category of smooth ($\infty $-differentiable)
manifolds with forgetful functor $f:{\cal M}\to Top$, which 
defines a local structure and the presheaves of these structures 
are sheaves.  On the category ${\cal M}$ there is the {\it 
tangent} functor $T:{\cal M}\to {\cal M}:M\mapsto T(M).$

Its iterations give us almost all interesting functors on ${\cal M}.$
Among them we shall note the following:

\begin{enumerate}
  \item[--]  The cotangent functor $T^{*}:{\cal M}\to {\cal 
M}:M\mapsto T^{*}(M).$ 

\item[--]  For a manifold $M$  and natural 
  number $k=0,1,\ldots $ the functor of $k-$jets $J^k:{\cal M}\to 
{\cal M}:N\mapsto J^k(M,N).$ 

\item[--]  For a manifold $M$, $x\in 
M,$ and natural number $k=0,1,\ldots $ the functor of $k-$jets at 
  the point $x$ $J_x^k:{\cal M}\to {\cal M}
J_x^k(M,N).$ \end{enumerate}

Any category ${\cal C}$ of structures on smooth manifolds
(or on ${\cal M}/$) has an additional structure, which give us a
possibility to define "smooth families of morphisms".

\begin{definition}
Let $M,M',M''\in {\cal M}.$ A map
\begin{equation*}
\Phi :M\to {\cal M}(M^{\prime },M^{\prime \prime }):x\mapsto \Phi _x
\end{equation*}
is called a  {smooth family} of morphisms if there exists a smooth map
$\phi :M\times M^{\prime }\to M^{\prime \prime }$ such that
\begin{equation*}
\forall x\in M,\ x^{\prime }\in M^{\prime }\quad \Phi _x(x^{\prime })
=\phi (x,x^{\prime }).
\end{equation*}
\end{definition}

Thus we get the class of categories with smooth families and it 
appears the natural condition on functors.

\begin{definition}
A functor is called a {smooth functor} if it maps each smooth family to
a smooth family.
\end{definition}

Of course all functors $T,T^*, J^k, J^k_x$ are smooth.

\subsection{Double Categories as additional structure \\
on categories}

In any category ${\cal C}$ with bundle products for some morphisms we can
define so-called intern categories. This is a monoid in the 
multiplicative category ${\cal C}//{O}$ of pairs of (special) 
morphisms $D, R: {M \to O}$ with the bundle product:

for $\xi  = (D, R: {M \to O})$ and $\xi' = (D', R': {M \to O})$ 
\, we get $\xi\star \xi'  = (D\circ \pi_1, R'\circ \pi_2:  
{M\times_O M' \to O})$ \,  where
the unit objects $\id_M: 0 \to M$ and $\id_{M'}: 0 \to M'$
and the following diagram is commutative
\begin{equation*}
\begin{CD}
  M\times_O M'  @>{\pi_2}>>  M'\\
  @V{\pi_1}VV  @VV{R'\, .}V  \\
  M  @>R>> O
\end{CD}
\end{equation*}

So an intern category is an object $\xi =(D, R: M ->O)$ with a
multiplication $\mu :\xi \star \xi' \to \xi$ and the unit $\id_M:O\to M$.

Now we consider such intern category as the category $\Cat$ of 
categories and will call it as {\it double categories} [20].

\begin{definition}
A double category $D$ consists of the following:

(1) A category $D_0$ of objects $Obj(D_0)$ and morphisms $Mor(D_0)$ of $0$%
-level.

(2) A category $D_1$ of morphisms $Obj(D_1)$ of $1$-level and morphisms $%
Mor(D_1)$ of $2$-level.

(3) Two functors $d,r:D_1\overrightarrow{\to }D_0.$

(4) A composition functor
\begin{equation*}
\ast :D_1\times _{D_0}D_1\to D_1
\end{equation*}
where the bundle product is defined by commutative diagram
\begin{equation*}
\begin{array}{ccc}
D_1\times _{D_0}D_1 & \underrightarrow{\pi _2} & D_1 \\
\pi _1\downarrow \quad  &  & \quad \downarrow d \\
D_1 & \underrightarrow{r} & D_0
\end{array}
\end{equation*}

(5) A unit functor $ID:D_0\to D_1$, which is a section of $d,r$.

\end{definition}

There are strong and weak double categories.

Now we see that for two objects $A,B\in Obj(D_0)$ there are 
$0$-level morphisms $D_0(A,B)$ which we note by ordinary arrows 
$f:A\to B,$ and $1$-level morphisms $D_{(1)}(A,B)$, which we 
note by the arrows $\xi :A\Rrightarrow B$ for $A=d(\xi )$ and 
$B=r(\xi )$. So with a $2$-level morphism $\alpha :\xi \to \xi 
^{\prime }$, where $\xi :A\Rrightarrow B$ and $ \xi ^{\prime 
}:A^{\prime }\Rrightarrow B^{\prime }$ we can associate the 
following diagram
\begin{equation*}
\begin{array}{ccccc}
A & \overset{\xi }{\Rrightarrow } & B &  & \xi \\
d(\alpha )\downarrow \qquad &  & \qquad \downarrow r(\alpha ) & \qquad
\longmapsto \qquad & \quad \downarrow \alpha \\
A^{\prime } & \overset{\xi ^{\prime }}{\Rrightarrow } & 
B^{\prime } &  & \xi^{\prime }
\end{array}
\end{equation*}

and arrow $\alpha :d(\alpha )\Rrightarrow r(\alpha )$.

On each level we have the corresponding compositions:
\begin{equation*}
\begin{array}{cccc}
\begin{array}{c}
\text{0-level} \\
\end{array}
&
\begin{array}{c}
(A\overset{f}{\rightarrow }B\overset{g}{\rightarrow }C) \\
\xi \overset{\alpha }{\rightarrow }\eta \overset{\beta }{\rightarrow }%
\varsigma
\end{array}
&
\begin{array}{c}
\mapsto \\
\mapsto
\end{array}
&
\begin{array}{c}
g\circ f:A\rightarrow C \\
\beta \circ \alpha :\xi \rightarrow \varsigma
\end{array}
\\
\text{1-level} & (A\overset{\xi }{\Rrightarrow }B\overset{\eta }{%
\Rrightarrow }C) & \mapsto & \eta *\xi :A\Rrightarrow C \\
\text{2-level} & (f\overset{\alpha }{\Rrightarrow }g\overset{\beta }{%
\Rrightarrow }h) & \mapsto & \beta *\alpha :f\Rrightarrow h
\end{array}
\end{equation*}

The composition on 2-level associated with the diagram
\begin{equation*}
\begin{array}{ccccc}
A & \overset{\xi }{\Rrightarrow } & B &  & \xi \\
d(\alpha )\downarrow \qquad &  & \qquad \downarrow r(\alpha ) &  & \quad
\downarrow \alpha \\
A^{\prime } & \overset{\xi ^{\prime }}{\Rrightarrow } & B^{\prime } & \qquad
\longmapsto \qquad & \xi ^{\prime } \\
d(\alpha ^{\prime })\downarrow \qquad \; &  & \qquad \;\downarrow r(\alpha
^{\prime }) &  & \quad \;\downarrow \alpha ^{\prime } \\
A^{\prime \prime } & \overset{\xi ^{\prime \prime }}{\Rrightarrow } &
B^{\prime \prime } &  & \xi ^{\prime \prime }
\end{array}
\end{equation*}

Thus, a double category $D$ consists of 

\begin{itemize}
\item  four sets $Obj(D_0),Mor(D_0),Obj(D_1),Mor(D_1),$ and eight maps of
type $d,r$%
\begin{equation*}
\begin{array}{ccc}
Obj(D_1) & \overleftarrow{\leftarrow } & Mor(D_1) \\
\downarrow \downarrow  &  & \downarrow \downarrow  \\
Obj(D_0) & \overleftarrow{\leftarrow } & Mor(D_0)
\end{array}
\end{equation*}

\item  two categories are associated $D_0,$ $D_1,$ and 
{\bf almost categories}: $D_{(2)}$ with the set of objects $Obj(D_0)$ 
and the set of morphisms $Obj(D_1),$ $D_{(3)}$ with the set 
of objects $Mor(D_0)$ and the set of morphisms $Mor(D_1),$

\item  $r,d:D_{(3)}\rightarrow D_{(2)}$ are almost functors.
\end{itemize}

Now we can define for double categories {\bf double (category) functors} and
their {\bf morphisms}, {\bf double subcategories} , the category $DCat$ 
of double categories, {\bf equivalence} of double categories, {\bf dual 
double categories} (changed direction of 1-level morphisms, i.e. $d,r$ 
are transposed), and so on.

\begin{definition}
A double category functor $F:D\rightarrow D^{\prime }$ is a pair $%
F_0:D_0\rightarrow D_0^{\prime },F_1:D_1\rightarrow D_1^{\prime }$ of usual
functors such that
\begin{eqnarray*}
d^{\prime }\circ F_1=F_0\circ d,\quad r^{\prime }\circ F_1=F_0\circ r, \\
\forall \;\xi ,\xi ^{\prime }\in Obj(D_1)\qquad \varphi _{\xi ,\xi ^{\prime
}}:F_1(\xi *\xi ^{\prime })\widetilde{\rightarrow }F_1(\xi )*^{\prime
}F_1(\xi ^{\prime }), \\
\forall \;A\in Obj(D_0)\qquad \varphi _A:F_1(ID_A)\widetilde{\rightarrow }%
ID_{F_0(A)}.
\end{eqnarray*}
\end{definition}

\subsection{Examples of Double Categories}

Examples considered bellow show that double categories are sufficiently
natural for mathematics.

\begin{example}
Bicategories are the partial case of double category $D$
when the category $D_0$ is trivial, i.e. has only identical morphisms and
composition of 1-level and 2-level morphisms are associative.
\end{example}

\begin{example}
For each category $C$ we have the canonical double category $Morph(C)$ of
morphisms. Let $C$ be a category, $T$ be the diagram $\bullet \to \bullet ,$
$TC$ be the category of diagrams in $C$ of type $T$, let $D_0=C$ and $D_1=TC.
$ The functor $d$ maps the diagram $f:A\to B$ into the object $A,$ the
functor $r$ maps this diagram into the object $B,$ and so on. It is easy to
see that we get a double category $D$ which is noted by $Morph(C)$. Here $%
Obj(D_1)=Mor(D_0)$ , a 2-level morphism $f\Rrightarrow g$ is a pair $(u,v)$
of morphisms $u,v\in Mor(C)$ with usual composition  from the commutative diagram
\begin{equation*}
\begin{array}{ccc}
A & \overset{u}{\rightarrow } & A^{\prime } \\
f\downarrow ~ &  & ~\downarrow f^{\prime } \\
B & \overset{v}{\rightarrow } & B^{\prime }
\end{array}
\end{equation*}

\end{example}

\begin{example}
Let $C$ be a category with bundle products, i.e. for all morphisms $u,v$ to $%
Y$ the universal square
\begin{equation*}
\begin{array}{ccc}
X\times _ZY & \longrightarrow  & Y \\
\downarrow  &  & \quad \downarrow v \\
X & \overset{u}{\longrightarrow } & Z
\end{array}
\end{equation*}
exists. And let $T$ be the following diagram
\begin{equation*}
\bullet \leftarrow \bullet \to \bullet ,
\end{equation*}
$TC$ be the category of diagrams in $C$ of type $T$. Now we define the
double category $D$ with $D_0=D$ and $D_1=TC.$ Two functors
\begin{equation*}
d,r:TC\to C,
\end{equation*}
where the functor $d$ maps the diagram $A\leftarrow M\rightarrow B$ into the
object $A,$ the functor $r$ maps this diagram into the object $B.$ The
composition: for two $1$-level morphisms $\xi =(A\overset{\pi }{\leftarrow }M%
\overset{f}{\rightarrow }B):A\Rrightarrow B$ and $\xi ^{\prime }=(B\overset{%
\pi ^{\prime }}{\leftarrow }M^{\prime ^{\prime }}\overset{f^{\prime }}{%
\rightarrow }C):B\Rrightarrow C$ we define their composition $\xi ^{\prime
}\circ \xi =(A\overset{\pi \circ \pi _1}{\leftarrow }M\times _BM^{\prime }%
\overset{f\circ \pi _2}{\rightarrow }C)$ where the bundle product is defined
by the universal diagram
\begin{equation*}
\begin{array}{ccc}
{{M\times _BM^{\prime }}} & \underrightarrow{\pi _2} & {M^{\prime }} \\
\pi _1\downarrow \quad  &  & \quad \downarrow \pi ^{\prime } \\
M & \overset{f}{\rightarrow } & B
\end{array}
\end{equation*}
A 2-level morphism is a triple $\alpha =(u,v,w):\xi \rightarrow \xi ^{\prime
}$ from the following commutative diagram
\begin{equation*}
\begin{array}{ccccc}
\underset{}{M} & \overset{f}{\rightarrow } & B &  &  \\
\underset{}{\overset{}{\pi \downarrow \ }} & \ \searrow v &  & \ \searrow w
&  \\
\underset{}{\overset{}{A}} &  & M^{\prime } & \overset{f^{\prime }}{%
\rightarrow } & B^{\prime } \\
& \underset{}{\overset{}{\ \searrow u}} & \pi ^{\prime }\downarrow \ \  &  &
\\
&  & \overset{}{\underset{}{A^{\prime }}} &  &
\end{array}
\end{equation*}
with the evident composition.
\end{example}

\begin{example}
Let us consider  a multiplicative (tensor) category $(C,\otimes 
,U,u)$.  Then we have the double category with 
$D_1=C,$ and $D_0=(*,*)$, e.c. a trivial category with one object 
and one morphism.  The composition is 
\begin{equation*} D_1\times 
_{D_0}D_1=C\times C\overset{\otimes }{\rightarrow }C.  
\end{equation*} 
Let us consider it in more details. Let $(C,\otimes ,U,u)$ be a 
multiplicative (tensor) category with multiplication 
\begin{equation*} 
\otimes :C\times C\rightarrow C:(X,Y)\mapsto 
X\otimes Y, 
\end{equation*} 
for the functor isomorphism of associativity 
\begin{equation*} 
\varphi :\otimes \circ 
(id,\otimes )\rightarrow \otimes \circ (\otimes ,id) 
\end{equation*} 
we write 
\begin{equation*} 
\varphi 
_{X,Y,Z}:X\otimes (Y\otimes Z)\rightarrow (X\otimes Y)\otimes Z 
\end{equation*}
so the pentagon is commutative

\begin{equation*}
{\small
\begin{array}{ccccc}
X\otimes (Y\otimes (V\otimes W)) & \overset{\varphi _{X,Y,V\otimes W}}{
\longrightarrow } & (X\otimes Y)\otimes (V\otimes W) & \overset{\varphi
_{X\otimes Y,V,W}}{\longrightarrow } & ((X\otimes Y)\otimes V)\otimes W \\
\underset{}{\overset{}{id_X\otimes \varphi _{Y,V,W}}}\downarrow \qquad \quad
\qquad \ \quad  &  &  &  & \varphi _{X,Y,V}\otimes id_W\uparrow \qquad \quad
\qquad \  \\
X\otimes ((Y\otimes V)\otimes W) &  & \overset{\varphi _{X,Y\otimes V,W}}{
\longrightarrow } &  & (X\otimes (Y\otimes V))\otimes W
\end{array}}
\end{equation*}

Then we have the double category $D$ with $D_0=C$ and $D_1$ such that
\begin{equation*}
Obj(D_1)=\{(X,x)|A,B,X\in Obj(C),\quad \/x:X\otimes A\rightarrow B\}.
\end{equation*}
So, we write $\xi =(X,x):A\Rrightarrow B$ and for $\xi \in Obj(D_1)$ we
denote $\xi =(X_\xi ,x_\xi ),$ $d(\xi )=A_\xi ,\quad r(\xi )=B_\xi .$
2-level morphisms
\begin{eqnarray*}
D_1(\xi ,\xi ^{\prime }) &\!\!\!=\!\!\!& 
\{(f_1,f_2,f_3)\ |\ commutative\;diagram\;
\begin{array}{ccc}
X\otimes A & \overset{x}{\longrightarrow } & B \\
f_3\otimes f_1\downarrow \qquad ~\ \  &  
& \quad \downarrow f_2^{\prime } \\
X^{\prime }\otimes A^{\prime } 
& \overset{x^{\prime }}{\longrightarrow } &
B^{\prime }
\end{array}
\}
\end{eqnarray*}
and $d(f_1,f_2,f_3)=f_1,\quad r(f_1,f_2,f_3)=f_2.$

Composition $D_1\times _{D_0}D_1\rightarrow D_1$ is defined as 
follows

for $A\overset{\xi }{\Rrightarrow }B\overset{\xi ^{\prime }}{\Rrightarrow }
B^{\prime }$ $\quad \xi \circ \xi ^{\prime }=(A,B^{\prime },X^{\prime
}X,x^{\prime \prime }),$ where $x^{\prime \prime }$ is the following
composition

$(X^{\prime }\otimes X)\otimes A\overset{\varphi _{X^{\prime },X,A}^{-1}}{
\longrightarrow }X^{\prime }\otimes (X\otimes A)\overset{id_{X^{\prime
}}\otimes x}{\longrightarrow }X^{\prime }\otimes B\overset{x^{\prime }}{
\longrightarrow }B^{\prime }.$

Associativity. For $A\overset{\xi }{\Rrightarrow }B\overset{\xi ^{\prime }}{
\Rrightarrow }B^{\prime }\overset{\xi ^{\prime \prime }}{\Rrightarrow }
B^{\prime \prime }$ the left column gives $x_{\xi ^{\prime \prime 
}\circ (\xi ^{\prime }\circ \xi )}$, the right column gives 
$x_{(\xi ^{\prime \prime }\circ \xi ^{\prime })\circ \xi }$

\begin{equation*}
\begin{array}{ccc}
(X^{\prime \prime }\otimes (X^{\prime }\otimes X))\otimes A &  & ((X^{\prime
\prime }\otimes X^{\prime })\otimes X)\otimes A \\
\underset{}{\overset{}{\varphi _{X^{\prime \prime },X^{\prime }\otimes
X,A}^{-1}\downarrow \qquad \quad \quad }} &  & \varphi _{X^{\prime \prime
}\otimes X^{\prime },X,A}^{-1}\downarrow \qquad \quad \quad  \\
X^{\prime \prime }\otimes ((X^{\prime }\otimes X)\otimes A) &  & (X^{\prime
\prime }\otimes X^{\prime })\otimes (X\otimes A) \\
\underset{}{\overset{}{id_{X^{\prime \prime }}\otimes \varphi _{X^{\prime
},X,A}^{-1}\downarrow \qquad \quad \qquad }} &  & id_{X^{\prime \prime
}\otimes X^{\prime }}\otimes x\downarrow \qquad \quad \quad  \\
X^{\prime \prime }\otimes (X^{\prime }\otimes (X\otimes A)) &  & (X^{\prime
\prime }\otimes X^{\prime })\otimes B \\
\underset{}{\overset{}{id_{X^{\prime \prime }}\otimes (id_{X^{\prime
}}\otimes x)\downarrow \qquad \quad \quad \qquad \quad }} &  & \varphi
_{X^{\prime \prime },X^{\prime },B}^{-1}\downarrow \qquad \quad  \\
X^{\prime \prime }\otimes (X^{\prime }\otimes B) & = & X^{\prime \prime
}\otimes (X^{\prime }\otimes B) \\
\underset{}{\overset{}{id_{X^{\prime \prime }}\otimes x^{\prime }\downarrow
\qquad \quad \quad }} &  & id_{X^{\prime \prime }}\otimes x^{\prime
}\downarrow \qquad \quad \quad  \\
X^{\prime \prime }\otimes B^{\prime \prime } & = & X^{\prime \prime }\otimes
B^{\prime } \\
x^{\prime \prime }\downarrow \quad  &  & x^{\prime \prime }\downarrow \quad
\\
B^{\prime \prime } &  & B^{\prime \prime } \\
&  &  \\
&  &
\end{array}
\end{equation*}
So we have isomorphism

$(\varphi _{X^{\prime \prime },X^{\prime },X},id_{A^{\prime
}},id_{B^{\prime }}):\xi ^{\prime \prime }\circ (\xi ^{\prime }\circ \xi
)\rightarrow (\xi ^{\prime \prime }\circ \xi ^{\prime })\circ \xi .$
\end{example}

\subsubsection{Bundle of Categories}

Let $\varphi :{\cal F\to C}$  be a functor and for all objects 
$U\in Obj{\cal  C}$ and we denote by ${\cal F}_U=\varphi 
^{{-1}}(U,id_U)$ the subcategory of $F$ with 
\begin{equation*} 
Obj\;{\cal F}_U=\left\{ {u\in Obj{\cal F\ }|\ \varphi (u)=U}\right\} ,
\end{equation*}
\begin{equation*}
Mor\;{\cal F}_U=\left\{ {f\in Mor\;{\cal F\ }|\ \varphi (f)=id_U}\right\} .
\end{equation*}
Let $(f:v\to u)\in Mor{\cal F},$ $\varphi (f:v\to u)=(g:V\to U).$ Then one
tells that  $f$ is {\it Descartes's morphism }, or that $v$ is {\it 
inverse image $g^{*}(u)$ of the object $u$}, if $\forall v^{\prime }\in Obj(
{\cal F}_V)$ the map
\begin{equation*}
f_{*}:{\cal F}_V(v^{\prime },v)\to {\cal F}_g(v^{\prime },u):h\mapsto f\circ
h
\end{equation*}
is a bijection. Here we have 
\begin{equation*}
{\cal F}_g(v,u)\overset{def}{=}\left\{ {h\in {\cal F}(v,u)\ |\ \varphi (h)=g}.
\right\}
\end{equation*}
So we have the diagram
\begin{equation*}
\begin{array}{ccc}
{{\forall \ v^{\prime }}} &  &  \\
{\;\ \ \ \ \downarrow ^h} & {{\qquad \searrow \;^{f\circ h}}} &  \\
\ \ \ {v} & \underset{f}{\longrightarrow } & {u} \\
&  &  \\
V & \overset{g}{\longrightarrow } & {U}
\end{array}
\end{equation*}
A functor $P:{\cal F\to C}$ is called a {\it bundle of categories} if
inverse images allows exist and a composition two Descartes morphism is
Descartes morphism too. Then $g^{*}$ may be transfered to functor ${\cal F}(U)\to
{\cal F}(V)$, and $(g_1\circ g_2)^{*}$ will be canonical isomorphic to $%
g_2^{*}\circ g_1^{*}.$

\begin{example}
 The projection
\begin{equation*}
\Pi _1:Mor(Top)\to Top:(f:X\to Y)\mapsto X
\end{equation*}
is a bundle of categories. For different structures on topological spaces it
is not always truth for the category of all morphisms, but may be truth for
a subcategory.
\end{example}
\begin{example}
 Let {\bf Sub} be a subcategory in $Mor(Man)$ consists from submersions.
Then projection
\begin{equation*}
\Pi _2:{\bf Sub}\to Man:(f:X\to Y)\mapsto Y
\end{equation*}
is a bundle of categories and for each morphism $h\in Man(B^{\prime},B)$ we
have the functor of {\bf inverse image}:
\begin{equation*}
h^{*}:{\bf Sub}_B\to {\bf Sub}_{B^{\prime }}:(f:M\to B)\mapsto (B^{\prime
}\times _BM\to B^{\prime }).
\end{equation*}
The set $\Gamma (\pi )$ of sections of an submersion $\pi :M\to B$ is the
set of morphisms ${\bf Sub}(id_B,\pi )$.
\end{example}
\begin{example}
 Let ${\bf Mod}$ be the category of pairs $(R,M)$ where $R$ is a ring and $%
M$ is a left $R$-module. Let ${\bf Rings}$ be the category of rings. Then
the functor
\begin{equation*}
Mod\to Rings:(R,M)\mapsto R
\end{equation*}
is a bundle of categories and for each morphism $h\in {\bf Ring}(R^{\prime
},R)$ we have the functor of {\bf inverse image}:
\begin{equation*}
h^{*}:R\text{{\bf -}}{\bf mod}\to R^{\prime }\text{{\bf -}}{\bf mod}
:M\mapsto R^{\prime }\otimes _RM.
\end{equation*}
\end{example}

\subsection{Fibers of Functor Morphisms}

The Grothendieck's definition of a fiber of a functor morphism is applicable
to morphisms of functors from any category to the category {\bf Sets} of sets.
Let $F,G:{\cal C}\to Set,$ and $\varphi :F\to G$ be their morphism. For each
object $S\in Obj({\cal C})$ and an element $\alpha \in G(S)$ the {\it fiber 
$\varphi _\alpha $ of $\varphi $ over} $\alpha $ is the following functor
\begin{equation*}
\varphi _\alpha :{\cal C}/S\to Sets:f\mapsto \varphi _\alpha (f),
\end{equation*}
where for a morphism $f:T\to S$
\begin{equation*}
\varphi _\alpha (f)=\left\{ {\beta \in F(T)\ |\ G(f)\circ \varphi _T(\beta
)=\alpha }\right\} .
\end{equation*}
So we have the following diagram
\begin{equation*}
\begin{array}{ccccc}
{\varphi ^\alpha (f)\subset } & {F(T)} &  & {F(S)} &  \\
& {\varphi _T\downarrow \;\;} &  & {\;\;\downarrow \varphi _S} &  \\
& {G(T)} & \overset{G(f)}{\longrightarrow } & {G(S)} & {\ni 
\alpha\, . } \end{array} \end{equation*}

\section{Multiplicative structures
on categories}

\subsection{Concepts and state of the art}

The prototype of a category is the category $\Sets$ of sets and functions.
The  prototype of a 2-category is the category $\Cat$ of small categories
and functors.
$\Cat$ has more structure on it then a simple category because we have
natural transformations between functors.
This can be viewed in the following way:
The extra structure implies that every morphism set $\Hom(C,D)$
in $\Cat$ is actually not only a set but a category itself where composition
and identities in $\Cat$ are compatible with this categorical structure
on the $\Hom$-sets (i.e. composition and identities are functorial with
respect to the structure on the $\Hom$-sets).
A general category with this kind of extra
structure is called a 2-category.

The definition of a 2-category can be put in a more general setting (which
will be convenient below) by using the language of enriched categories.
A category ${\cal C}$ is enriched over a category ${\cal V}$ if every
$\Hom$-set in ${\cal C}$ has the structure  of an object in ${\cal V}$
and if composition and identities in ${\cal C}$ are compatible with
this extra structure on the $\Hom$-sets. So, a 2-category is a category
enriched over $\Cat$. Now, the (small) 2-categories again form a category
{\bf 2-Cat} and a 3-category can be defined as a category 
enriched over {\bf 2-Cat} (indeed, {\bf 2-Cat} turns out to be a 
3-category itself). In this way we can proceed iteratively to 
define $n$-categories and then $\omega$-categories as categories 
involving $n$-categorical structures of all levels.

A concrete recipe obtaining of monoidal (braided etc) 
2-categories via Hopf categories is proposed by Crane and Frenkel 
[27].  Namely, that it is supposed the 2-category of 
module-categories over a Hopf category now  plays an important 
role in 4-dimensional topology and TQFT.  Although the theory of 
Hopf categories is devised, in general, by Neuchl [28], 
interesting examples are still missing. In particular the Hopf 
category, underlying the Lusztig's canonical basis [29] of a 
quantized universal enveloping algebra, is not constructed yet.  
We propose to define it as a family of abelian categories of 
perverse $l$-adic sheaves equipped with some functors of 
multiplication and comultiplication [30]. These perverse 
sheaves are equivariant in the sense of Bernstein and Lunts 
[31].

It turns out that the notions of $n$-category and $\omega$-category are
not general enough for several interesting applications.
What one gets there are weak versions of these concepts (instead of
weak $n$-category sometimes the notions
bicategory, tricategory, etc. are used). Let us shortly explain what this
means: In a category it does not make sense to ask for equality of objects
but the appropriate notion is isomorphism. In the same way, in a 2-category
we should not ask for equality of morphisms but only for equality up to
an invertible 2-morphism (the morphisms between the morphisms, e. g. the
natural transformations in $\Cat$).
Applying this to the categorical structure itself (i.e. requiring
associativity and identity properties only up to natural equivalence)
leads to the notion of weak 2-category (or bicategory).
In the same way, we can weaken the structure of an $n$-category up to
the $(n-1)$-th level to obtain a weak $n$-category.

The point making this weakening an involved matter is that in general
we need so called coherence conditions in
addition to the weakened laws in order to assure that some properties,
known  from the strict case, hold. E.g., to assure that associativity
is iteratively applicable (i.e. that we can up to a $2$-isomorphism
rebracket composites involving more than three factors), we need a
coherence condition stating that even four factors can be rebracketed
(and the other cases follow then).
See the literature given above for the details.

A satisfactory version of a weak $n$-category for higher $n$ and of a weak
$\omega $-category was not available for a long time but now there are
several approaches at hand [32--34].
The relationship between these 
approaches and a universal understanding of these structures has 
still to be achieved.

\subsection{Multiplicative Categories}

\begin{definition}
A {\bf multiplication} in the category ${\cal C}$ is an associative functor
$$
\ast :{\cal C}\times {\cal C}\to {\cal C}:(X,Y)\mapsto X*Y.
$$
An {\bf associativity morphism} for $*$ is a functor isomorphism
$$
\varphi _{X,Y,Z}:X*(Y*Z)\to (X*Y)*Z
$$
such  that for any four objects $X,Y,Z,T$ the following diagram 
is commutative:

$$
\begin{array}{ccccc}
{X*(Y*(Z*T))} & \overset{\varphi _{X,Y,Z*T}}{\longrightarrow } & 
{(X*Y)*(Z*T)
} & \overset{\varphi _{X*Y,Z,T}}{\longrightarrow } & {X*Y*Z*T} \\
{{\quad \quad \quad \;\downarrow id_X*\varphi _{Y,Z,T}}} &  &  &  & 
{{}{}{}{
\quad \quad \quad \;\uparrow \varphi _{X,Y,Z}*id_T}} \\
{X*((Y*Z)*T)} &  & \overset{\varphi _{X,Y*Z,T}}{\longrightarrow } &  & {
(X*(Y*Z))*T}
\end{array}
$$
$$
\begin{CD}
{X*(Y*(Z*T))} @>{\varphi _{X,Y,Z*T}}>> {(X*Y)*(Z*T)}
@>{\varphi _{X*Y,Z,T}}>> {X*Y*Z*T} \\
@V{id_X*\varphi _{Y,Z,T}}VV &  &
@AA{\varphi _{X,Y,Z}*id_T}A \\
{X*((Y*Z)*T)} & @>{\varphi _{X,Y*Z,T}}>> & {(X*(Y*Z))*T}
\end{CD}
$$
An {\bf commutativity morphism} for $*$ is a functor isomorphism
$$
\psi _{X,Y}:X*Y\to Y*X
$$
such that for any two objects $X,Y$ we have
$$
\varphi _{X,Y}\circ \varphi _{Y,X}=id_{X*Y}:X*Y\to X*Y.
$$
Morphisms associativity $\varphi $ and commutativity $\psi $ are
{compatible} if for any three objects $X,Y,Z$ the following diagram is
commutative:
$$
\begin{array}{ccccc}
{X*(Y*Z)} & \overset{\varphi _{X,Y,Z}}{\longrightarrow } & {(X*Y)*Z} &
\overset{\psi _{X*Y,Z}}{\longrightarrow } & {Z*(X*Y)} \\
{{\quad \quad \quad \;\downarrow id_X*\psi _{Y,Z}}} &  &  &  & {{}{}{}{\quad
\quad \quad \;\uparrow \varphi _{Z,X,Y}}} \\
{X*(Z*Y)} & \overset{\varphi _{X,Z,Y}}{\longrightarrow } & {(X*Z)*Y} &
\overset{\psi _{X,Z}*id_Y}{\longrightarrow } & {(Z*X)*Y}
\end{array}
$$
$$
\begin{CD}
{X*(Y*Z)} & \overset{\varphi _{X,Y,Z}}{\longrightarrow } & {(X*Y)*Z} &
\overset{\psi _{X*Y,Z}}{\longrightarrow } & {Z*(X*Y)} \\
{{\quad \quad \quad \;\downarrow id_X*\psi _{Y,Z}}} &  &  &  & {{}{}{}{\quad
\quad \quad \;\uparrow \varphi _{Z,X,Y}}} \\
{X*(Z*Y)} & \overset{\varphi _{X,Z,Y}}{\longrightarrow } & {(X*Z)*Y} &
\overset{\psi _{X,Z}*id_Y}{\longrightarrow } & {(Z*X)*Y}
\end{CD}
$$
A pair $(U,u)$ where $U\in Obj({\cal C})$ and an isomorphism $u:U\to U*U$ is
called a {\bf unit object} for ${\cal C},*$ if the functor
$$
X\mapsto U*X:{\cal C}\to {\cal C}
$$
is equivalence of categories.
\end{definition}

\begin{definition}
A {\bf multiplicative category} is a collection $({\cal C},*,\varphi ,\psi
,U,u).$
\end{definition}

If there are some additional structures on category, then it is  
usually assumed that product $*$ and others elements of the 
collection are compatible with these structures.

\subsection{{${\cal C}$}-monoids or multiplicative objects.\\
 Monoidal categories and Monoids. Comonoids}

Let ${\bf C}=({\cal C},*,\varphi ,\psi ,U,u)$ be a multiplicative category.
An {\bf multiplicative object} in ${\bf C}$ or {\bf {${\cal C}$}-monoid} is
an object $M\in Obj({\cal C})$ with multiplication $\mu :M*M\to
M:(m,m^{\prime})\mapsto \mu(m,m^{\prime})$ and an unit $\varepsilon :U\to M$
such that the following axioms are faithful:

\begin{itemize}
\item[(1)]  Associativity: the following diagram is commutative
$$
\begin{array}{ccccc}
{M*(M*M)} &  & \overset{\varphi _{M,M,M}}{\longrightarrow } &  & {(M*M)*M}
\\
{\quad \quad \downarrow id_M*\mu } &  &  &  & {{}{}{}{\quad \quad \downarrow
\mu *id_M}} \\
{M*M} & \overset{\mu }{\longrightarrow } & {M} & \overset{\mu }{
\longleftarrow } & {M*M}
\end{array}
$$

\item[(2)]  Unit: the following diagram is commutative
$$
\begin{array}{ccccc}
{M} & \longrightarrow  & {U*M} & \overset{\psi _{U*M}}{\longrightarrow } & {
M*U} \\
{||} &  & {{}{\quad \qquad \downarrow \varepsilon *id_M}} &  & {{}{\quad
\qquad \downarrow id_M*\varepsilon }} \\
{M} & \overset{\mu }{\longrightarrow } & {M*M} & {=} & {M*M}
\end{array}
$$

\end{itemize}


\begin{example}
 Let $R$ be a commutative ring. The category $R${\bf -}${\bf mod}$ of $R$
-modules is a multiplicative category under the tensor product $\otimes _R$
with the unit object is the left $R$-module $R$. Multiplicative objects in
the category is $R-$algebras with units.
\end{example}

\begin{example}
 A small multiplicative category ${\cal C}$ is a multiplicative object of
the multiplicative category $Sets //Obj({\cal C})$.
\end{example}



 Multiplicative structures may be described in categories as
 monoids in a monoidal category.

 A monoidal category $({\cal C}, \otimes , K, \varphi, \ldots )$
consists of:

 $\otimes: {\cal C} \times {\cal C} \to {\cal C}$,
   $K\in \mbox{Ob} {\cal C}$  --  the unit object,

 and the functor-isomorphisms:

 $$
   \varphi_{A,B,C}:(A \otimes B) \otimes C \to  A \otimes (B \otimes C)
 $$

 $\psi_A :A\otimes K \to A$, $\ldots $,

 where $\otimes $  is symmetrical, if there exists  a 
functor-isomorphism $$ \theta_{A,B}:A \otimes B \to  B \otimes A. 
 $$

 A monoid in a monoidal category  $({\cal C}, \otimes , K, \varphi, \ldots )$
 is an object $M$ endowed a multiplication
 $$
   \mu: M\otimes M \to M
 $$
 and the unit morphism   $    \varepsilon: K \to M  $
 +  Axioms.

 A comonoid is a monoid in  $({\cal C}^{op}, \otimes , K, \varphi, \ldots )$.
 In ${\cal C}$ we have the comultiplication
 $$
   \Delta : M \to M\otimes M
 $$
 the counit   $ \eta: M \to K  $     +  Axioms.

 An action of a monoid $M$ on $A$ is defined by
 $$
   \alpha :M\otimes A \to A
 $$
   +  Axioms.

 A monoidal functor (a morphism of monoidal categories)
 of two monoidal categories is defined by
 $F:({\cal C}, \otimes , K) \to ({\cal C}', \otimes', K' ) $
 if
 $$
   F(A\otimes B) \cong  F(A) \otimes' F(B)
 $$
 and $F(K)\cong K'$.

 \begin{example}
 A monoidal category is a monoid in the monoidal category
 $({\cal C}at, \times)$ of categories with Cartesian product.
 \end{example}

 \begin{example}
 The category ${\cal S}ymm$ with objects $[n]$ for $n=0,1, \ldots$
 and morphisms
 $$
    {\cal S}ymm([n],[m])=
    \begin{cases}
     \emptyset , & \mbox{ if } n\neq m, \cr
          \Sigma _n , & \mbox{ if } n = m.
    \end{cases}
 $$
 where $ \Sigma _n$ is the group of permutations of $(1, \ldots , n)$.
 with the multiplication
 $$
   *: {\cal S}ymm \times {\cal S}ymm  \to {\cal S}ymm
 $$
 such that $[n]*[m] \cong [n+m-1]$ with the folowing identification
 of the inputs
 $$
   (1, \ldots , n)* (\overline{1}, \ldots , \overline{m})
   = (1, \ldots , n,\overline{2}, \ldots , \overline{m})
 $$
 which explanes the action of $*$ on morphisms.
 \end{example}

 \begin{example}
 Let
 $({\cal C}, \otimes , K)$ and $({\cal C}', \otimes', K' )$
 be two monoidal categories, $F \in \mbox{Ob}({\cal C C}')$
 and $F(K)=K'$.
 Then for such functors $F$ on the category  there is a monoidal
 structure and a monoid is defined by a functor morphism
  $$
   \mu_{A,B}:F(A)\otimes' F(B) \to F(A\otimes B)
 $$
 with natural axioms associativity and unit.
 \end{example}

 {\sl EXAMPLES 3.6--3.7} Bialgebras and Dual construction:

 {\it
 Algebras   as monoids in $k$-{bf vect}, $k$-{\bf alg},
 Bialgebras as comonoids in $k$-{\bf alg}, $k$-{\bf bialg}.

 Double Categories as monoids in the category of pairs of 
 functors.}

\section{ Categories of information transformers} 


\subsection{ Common structure of classes of information \\
 transformers}


It is natural to assume that for any information transformer $a$ there are
defined a couple of spaces: $\A$ and $\B$, the space of ``inputs'' (or input
signals) and the space of ``outputs'' (results of measurement, transformation,
processing, etc.). We will say that $a$ ``acts'' from $\A$ to $\B$ and denote
this as $a\:\A\>\B$. It is important to note that typically an information
transformer not only transforms signals, but also introduces some ``noise''. In
this case it is \NW{nondeterministic} and cannot be represented just by a mapping
from $\A$ to $\B$.

It is natural to study information transformers of similar type by
aggregating them into families endowed by a fairly rich algebraic
structure
 [5,11].
 Specifically, it is natural to assume that families of ITs
poses the following properties:

(a)  If $a\:\A\>\B$ and $b\:\B\>\C$ are two ITs, then their
\NW{composition} $b\.a\:\A\>\C$ is defined.

(b)  This operation of composition is \NW{associative}.

(c)  There are certain \NW{neutral} elements in these families, i.e., ITs that do not
introduce any alterations. Namely, for any space $\B$ there exist a
corresponding IT $i\<{\B}\: \B\>\B$ such that $i\<{\B}\.a = a$ and
$b\.i\<{\B} = b$.

Algebraic structures of this type are called \NW{categories}
[6,~8].

 Furthermore, we will assume, that to every pair of information
transformers, acting from the same space $D$ to spaces $\A$ and $\B$
respectively, there corresponds a certain IT $a*b$ (called \NW{product} of $a$ and
$b$) from $D$ to $\A\#\B$. This IT in a certain sense ``represents'' both ITs
$a$ and $b$ simultaneously. Specifically, ITs $a$ and $b$ can be
``extracted'' from $a*b$ by means of projections $\pi\<{\A,\B}$ and $\nu\<{\A,\B}$
from $\A\#\B$ to $\A$ and $\B$, respectively, i.e.,
$\pi\<{\A,\B}\.(a*b) = a$,  $\nu\<{\A,\B}\.(a*b) = b$. Note, that
typically, an IT $c$ such that $\pi\<{\A,\B}\.c = a$,  $\nu\<{\A,\B}\.c = b$ is not
unique, i.e., a category of ITs does not have products (in category-theoretic
sense
 [6--9]). Thus, the notion of a category of ITs demands for an accurate
formalization.

Analysis of classes of information transformers studied in [5, 10--18], 
gives grounds to consider these classes as categories that satisfy certain fairly general conditions.

\subsection{ Elementary axioms for categories of information transformers} %

In this subsection we set forward the main properties of categories of ITs. All the following study
will rely exactly on these properties.

In [5, 10--18]
it is shown (see also examples in section~8 below) that classes of information transformers can be
considered as morphisms in certain categories. As a rule, such categories do not have products,
which is a peculiar expression of nondeterministic nature of ITs in these categories. However, it
turns out that deterministic information transformers, which are usually determined in a natural way
in any category of ITs, form a subcategory with products. This point makes it possible to define a
``product'' of objects in a category of ITs. Moreover, it provides an axiomatic way to describe an
extension of the product operation from the subcategory of deterministic ITs to the whole category
of ITs.

\begin{definition} 
We shall say that a category $\CC$ is a \NW{category of information transformers} if the following
axioms hold:
\begin{enumerate}
\item 
There is a fixed subcategory of \NW{deterministic} ITs $\DD$ that contains all the objects of the
category $\CC$ ($\Ob(\DD)=\Ob(\CC)$).
\item
The classes of \NW{isomorphisms} in $\DD$ and in $\CC$ coincide, that is, all the isomorphisms in
$\CC$ are deterministic.
\item
The categories $\DD$ and $\CC$ have a common \NW{terminal object} $\Z$.
\item
The category $\DD$ has pairwise \NW{products}.
\item
There is a specified \NW{extension of morphism product} from the subcategory $\DD$ to the whole
category $\CC$, that is, for any object $\D$ and for any pair of morphisms $a\:\D\>\A$ and
$b\:\D\>\B$ in $\CC$ there is certain information transformer $a*b\:\D\>\A\#\B$ (which is also
called a \NW{product} of ITs $a$ and $b$) such that
\[
  \pi\<{\A,\B}\.(a*b)=a, \qquad
  \nu\<{\A,\B}\.(a*b)=b.
\]
\item 
Let $a\:\A\>\C$ and $b\:\B\>\D$ are arbitrary ITs in $\CC$, then the IT $a\*b$ defined by
Eq.{}~\rf{3} satisfy Eq.{}~\rf{5}:
\[
  (a\*b)\.(c*d) = (a\.c)*(b\.d).
\]
\item 
Equality{}~\rf{7} holds not only in $\DD$ but in $\CC$ as well, that is, \NW{product} of
information transformers is  ``\NW{commutative} up to isomorphism.''
\item 
Equality{}~\rf{9} also holds in $\CC$. In other words, \NW{product} of information transformers
is ``\NW{associative} up to isomorphism'' too.
\end{enumerate}
\end{definition}

Now let us make several comments concerning the above definition.

We stress that in the description of the extension of morphism product from the category $\DD$ to
$\CC$ (cf.~5.) we {\em do not require the uniqueness} of an IT $c\:\D\>\A\#\B$ that satisfy
conditions{}~\rf{1}.

Nevertheless, it is easily verified, that the equations{}~\rf{4} are valid for $c=a\*b$ not only in the
category $\DD$, but in $\CC$ as well, that is,
\[
  \pi\<{\C,\D}\.(a\*b) = a\.\pi\<{\A,\B}, \qquad
  \nu\<{\C,\D}\.(a\*b) = b\.\nu\<{\A,\B}.
\]
However, the IT $c$ that satisfy the equations{}~\rf{4} may be not unique. Note also that in the
category $\CC$ Eq.{}~\rf{2} in general does not hold.

Further, note that the axiom{}~%
6
immediately implies
\[
  (a\*b)\.(c\*d) = (a\.c)\*(b\.d).
\]

Finally note that any category that has a terminal object and pairwise products can be considered as a
category of ITs in which all information transformers are deterministic.

\section{ Category of information transformers as \\
 a monoidal category}

As we have already mentioned above in a category of ITs there are certain
``meaningful''   operations of product for objects and for morphisms.
However, these operations are not product operations in category-theoretic
sense. Nevertheless, every category of ITs is a \NW{monoidal category} (see,
e.g.,
 [6, 8]).

 First note, that every category $\DD$ with pairwise products and with
terminal object $\Z$ constitutes a monoidal category
$\TPL{\DD,\#,\Z,\al,\la,\rh}$, where $\#\:\DD\#\DD\>\DD$ is the
product functor and $\al\<{\A,\B,\C}\:(\A\#\B)\#\C\>\A\#(\B\#\C)$,
$\la\<{\A}\:\Z\#\A\>\A$, and $\rh\<{\A}\:\A\#\Z\>\A$ are the
obvious natural equivalences. Besides, as a category with
products, the category $\DD$ has a natural equivalence $\si$,
$\si\<{\A,\B}\:\A\#\B\>\B\#\A$, which interchanges components in a
product.

\begin{definition}
We will say that a category $\CC$ is a category of
information transformers over a subcategory (of deterministic ITs) $\DD$ if
the following three axioms hold.
\end{definition}

{\bf Axiom~1.} $\TPL{\CC,\#,\Z,\al,\la,\rh}$ is a monoidal
category for a certain: functor $\#\:\CC\#\CC\>\CC$, object $\Z$, and
natural equivalences $\al$, $\la$ and $\rh$.

 We will refer to morphisms of the category $\CC$ as \NW{information
transformers}.

{\bf Axiom~2.} The category $\CC$ has a subcategory $\DD$, such that all
the objects of $\CC$ are contained in $\DD$, $\Z$ is a terminal object in
$\DD$, and the functor $\#$ is a product functor on $\DD$.

Morphisms of the subcategory $\DD$ will be called
\NW{deterministic} information transformers.

 Thus, the following properties hold in the subcategory $\DD$:

(a)     There are natural transformations defined in $\DD$,
$\pi\<{\A,\B}\:\A\#\B\>\A$ and $\nu\<{\A,\B}\:\A\#\B\>\B$ that specify
projections on components of a product.

(b)     For any deterministic Its (morphisms in $\DD$) $a\:\C\>\A$ and
$b\:\C\>\B$ there exists e unique IT $c = a*b\:\C\>\A\#\B$ for which
$\pi\<{\A,\B}\.c = a$ and $\nu\<{\A,\B}\.c = b$;

(c)     $\DD$ is also a monoidal category with the natural equivalences
$\al$, $\la$ and $\rh$ explicitly expressed through $\pi$ and $\nu$, i.e.,
\[
 \la\<{\A} \eqdef \pi\<{\Z,\A},
 \qquad
 \rh\<{\A} \eqdef \nu\<{\A,\Z},
\]
\[
 \al\<{\A,\B,\C} \eqdef
 (\pi\<{\A,\B}\.\pi\<{\A\#\B,\C})*
 \Bigl((\nu\<{\A,\B}\.\pi\<{\A\#\B,\C})*\nu\<{\A\#\B,\C}\Bigr).
\]

(d)     There is a natural equivalence of ``object transposition'' $\si$ defined
on $\DD$:
\[
 \si\<{\A,\B} \eqdef \nu\<{\A,\B}*\pi\<{\A,\B}\:\A\#\B\>\B\#\A.
\]

(e)     There is a ``diagonal'' natural transformation $\de$ defined on $\DD$:
\[
 \de\<{\C} \eqdef i\<{\C}*i\<{\C}\:\C\>\C\#\C.
\]

 Note that with the help of the ``diagonal'' natural transformation the
product of morphisms $a*b$ may be expressed through their ``functorial
product'' $a\#b$, i.e., $a*b = (a\#b)\.\de\<{\C}$.

 Let us stress here, that we do not require that $\de$ is a natural
transformation on the whole category $\CC$. Furthermore, typically, in
many important examples of categories of ITs $\de$ is not a natural
transformation. Such categories do not have products in category-theoretic
sense. However we can extend the product operation for morphisms from
the subcategory $\DD$ to $\CC$. Specifically, we define in $\CC$:
\[
 a*b \eqdef (a\#b)\.\de\<{\C}.
\]

{\bf Axiom 3.} Natural transformations $\pi$, $\nu$ and $\si$ (in the category
$\DD$) are natural transformations in the whole category $\CC$ as well.

\begin{trm}1
Definitions 4.1 and 5.1 are equivalent.
\end{trm}

\section{ IT-Category as Kleisli category}


\subsection{ Concept of distribution. Kleisli category}


 The two equivalent definitions presented above provide the minimal
conceptual background for studying categories of ITs, e.g., for definition and
analysis of informativeness, semantic informativeness, decision problems,
etc. [1--5, 10--18].
However these definitions do not provide any tools for
constructing categories of ITs on the basis of more elementary concepts. The
concept of \NW{distribution} is one of the most important and it plays a critical
role in the uniform construction of a wide spectrum of IT-categories. Its
importance is connected to the observation that in many important
IT-categories an information transformer $a\:\A\>\B$ may be represented by
a morphisms from $\A$ to the ``object of distributions'' over $\B$. For
example, a probabilistic transition distribution (an IT in the category of
stochastic ITs) may be represented by a certain measurable mapping $\A$ to
the space of distributions on $\B$.

 Thus, we will suppose that on some fixed ``base'' category $\DD$
(category of deterministic ITs) there defined a functor $T$, which
takes an object $\A$ to the object $T\A$ of ``distributions'' on
$\A$. Besides, we assume that there are two natural
transformations connected to this functor: $\et\:I\>T$ and
$\mu\:TT\>T$. Informally, $\et\<{\A}\:\A\>T\A$ takes an element of
$\A$ to a ``discrete distribution, concentrated on this element'',
and $\mu\<{\A}\:TT\A\>T\A$ ``mixes'' (averages) a distribution of
distributions on $\A$, by transforming it to a certain
distribution on $\A$. Besides, there are natural ``coherence''
conditions for $\et$ and $\mu$:
\[
 \mu\<{\A} \. T\mu\<{\A}
 =
 \mu\<{\A} \. \mu\<{T\A}
\]
and
\[
 \mu\<{\A} \. T\et\<{\A} = i\<{T\A}
 \qquad
 \mu\<{\A} \. \et\<{T\A} = i\<{T\A}
\]
that may be presented by the
following commutative diagrams:
\[
\Dia(30,20){
\Put(0,0){TT\A}
\Put(30,0){T\A}
\Put(0,20){TTT\A}
\Put(30,20){TT\A}
\Vec(6,0;1,0;20)\Put(15,3){\mu}
\Vec(0,17;0,-1;13)\Put(4,10){\mu T}
\Vec(8,20;1,0;16)\Put(15,17){T\mu}
\Vec(30,17;0,-1;13)\Put(27,10){\mu}
}
\qquad\qquad
\Dia(60,20){
\Put(0,20){T\A}
\Put(30,20){TT\A}
\Put(60,20){T\A}
\Put(30,0){T\A}
\Vec(4,17;3,-2;22)\Put(10,10){T}
\Vec(5,20;1,0;19)\Put(15,17){\et T}
\Vec(30,17;0,-1;13)\Put(27,10){\mu}
\Vec(55,20;-1,0;19)\Put(45,17){T\et}
\Vec(56,17;-3,-2;22)\Put(50,10){T}
}
\]

Commutativity of the square means that for any ``third-order
distribution'' on $\A$ (i.e. distribution on a collection of
distributions on a family of distributions on $\A$) the result of
``mixing'' of distributions does not depend on the order of
``mixing''. More precisely, the result of mixing over the ``top''
(third order, element of $TTT\A$) distribution first and mixing
the resulting second-order distribution next should give the same
result as for mixing over ``intermediate'' (second-order, elements
of $TT\A$) distributions first and then mixing the resulting
second order distribution. Commutativity of the left triangle
means that mixing of a second order distribution, concentrated in
one element (which is itself a distribution on $\A$) gives this
distribution. Finally, commutativity of the right triangle means
if we take some distribution on $\A$, transform it to ``the same''
distribution of singletons and then mix the resulting second-order
distribution, we will obtain the original distribution.

 It is well known, that a collection $\TPL{T, \et, \mu}$, satisfying the two
commutative diagrams above, is called a \NW{triple} (monad) 
 [6, 8, 23]
on the category $\DD$.

 The concept of triple provides an elegant technique of constructing a
category of ITs $\CC$ on the basis of the category of deterministic ITs, as a
\NW{Kleisli} category
 [6, 23].
In this construction each morphisms $a\:\A\>\B$ in
the category $\CC$ is determined by a morphism $a'\:\A\>T\B$ of the
category $\DD$. The composition $a\.b$ of ITs $a\:\A\>\B$ and $b\:\B\>\C$ in
$\CC$ is represented by the morphism
$$
 (b\.a)' \eqdef \mu\<{\C}\.Tb'\.a'
$$
\[
\Dia(80,3){
\Put(10,0){(b\.a)'\;=\;\A}
\Vec(22,0;1,0;14)\Put(30,3){a'}
\Put(40,0){T\B}
\Vec(44,0;1,0;11)\Put(50,3){Tb'}
\Put(60,0){TT\C}
\Vec(65,0;1,0;11)\Put(70,3){\mu}
\Put(80,0){T\C}
}
\]
in $\DD$, and any deterministic IT $c\:\C\>\D$ (in $\CC$) are determined by
the morphism
$$
 c' \eqdef \et\<{\D}\.c
$$
\[
\Dia(50,3){
\Put(5,0){c'\;=\;\C}
\Vec(13,0;1,0;14)\Put(20,3){c}
\Put(30,0){\D}
\Vec(33,0;1,0;13)\Put(40,3){\et}
\Put(50,0){T\D}
}
\]
in $\DD$.

\subsection{ Independent distribution.\\
 Monoidal Kleisli category}


 The main factor in the construction of the category of ITs as a Kleisli
category is equipping it with a structure of monoidal category. For this
purpose we introduce a natural transformation $\ga\:\#T\>T\#$,
$\ga\<{\A,\B}\:T\A\#T\B\>T(\A\#\B)$, which ``takes'' a pair of distributions to
their ``independent joint distribution'' (see also
 [35]).
 Then the product
$c = a*b$ of ITs $a\:\D\>\A$ and $b\:\D\>\B$ (in $\CC$) is determined by
the morphism
$$
 c' \eqdef \ga\<{\A,\B}\.(a'*b')
$$
\[
\Dia(60,30){
\Put(0,15){\D}
\Put(30,15){T\A\#T\B}
\Vec(4,15;1,0;17)\Put(15,17){a'{*}b'}
\Put(60,15){T(\A\#\B)}
\Put(30,30){T\A}
\Vec(4,17;2,1;22)\Put(14,26){a'}
\Vec(30,18;0,1;9)\Put(34,22){\pi T}
\Put(30,0){T\B}
\Vec(4,13;2,-1;22)\Put(14,4){b'}
\Vec(30,12;0,-1;9)\Put(34,8){\nu T}
\Vec(40,15;1,0;11)\Put(45,18){\ga}
}
\]
in $\DD$. Note, that $a'*b'$ here exists and is uniquely defined
since $\DD$ is a category with products.

\begin{trm}2
Suppose that $\DD$ is a category with pairwise products and with
terminal object $\Z$; $\pi$, $\nu$, $\al$, $\si$ are the
corresponding natural transformations, and $\TPL{T, \et, \mu}$ is
a triple on $\DD$ with $\et\<\B$ monomorphic for every $\B$. Then
the generated Kleisli category $\CC$, equipped with a natural
transformation $\ga$, is a category of information transformers if
and only if the following compatibility conditions of $\ga$ with
the natural transformations $\pi$, $\nu$, $\al$, $\si$, $\et$, and
$\mu$ hold:

\medskip
\noindent{\boldmath
$\pi$-$\ga$ and $\nu$-$\ga$ conditions:}
\[
 T\pi\<{\A,\B} \. \ga\<{\A,\B} = \pi\<{T\A,T\B}
 \qquad
 T\nu\<{\A,\B} \. \ga\<{\A,\B} = \nu\<{T\A,T\B}
\]

\noindent{\boldmath
$\si$-$\ga$ condition:}
\[
 T\si\<{\A,\B} \. \ga\<{\A,\B} = \ga\<{\B,\A} \. \si\<{T\A,T\B}
\]

\noindent{\boldmath
$\al$-$\ga$ condition:}
\[
 T\al\<{\A,\B,\C} \. \ga\<{\A\#\B,\C} \. (\ga\<{\A,\B}\#i\<{T\C})
 =
 \ga\<{\A,\B\#\C} \. (i\<{T\A}\#\ga\<{\B,\C}) \. \al\<{T\A,T\B,T\C}
\]

\noindent{\boldmath
$\mu$-$\ga$ condition:}
\[
 \mu\<{\A\#\B} \. T\ga\<{\A,\B} \. \ga\<{T\A,T\B}
 =
 \ga\<{\A,\B} \. (\mu\<{\A}\#\mu\<{\B})
\]

\noindent{\boldmath
$\et$-$\ga$ condition:}
\[
 \ga\<{\A,\B} \. (\et\<{\A}\#\et\<{\B}) = \et\<{\A\#\B}
\]

\end{trm}

Thus, construction of a categories of ITs is, in effect, reduced to
selection of a base category $\DD$, a functor $T\:\DD\>\DD$, and a natural
transformation $\ga\:\#T\>T\#$.

All these conditions have rather transparent meaning that we will
try to comment below.

For better understanding we also provide the
corresponding commutative diagrams in which we omit the obvious
indices for the sake of readability:

{\boldmath
$\pi$-$\ga$ and $\nu$-$\ga$ conditions.}
Marginal distributions extracted from independent joint
distribution coincide with the original distributions:
\[
\Dia(60,30){
\Put(0,15){T\A\#T\B}
\Put(60,15){T(\A\#\B)}
\Vec(10,15;1,0;40)\Put(30,13){\ga}
\Put(30,30){T\A}
\Vec(6,18;2,1;20)\Put(14,26){\pi T}
\Vec(54,18;-2,1;20)\Put(46,26){T\pi}
\Put(30,0){T\B}
\Vec(6,12;2,-1;20)\Put(14,4){\nu T}
\Vec(54,12;-2,-1;20)\Put(46,4){T\nu}
}
\]

{\boldmath $\si$-$\ga$ condition.} Transposition of
components of an independent joint distribution leads to the
corresponding transformation of the joint distribution, i.e.,
Independent joint distribution is ``invariant'' with respect to
transposition of its components. More precisely, we can say that
the independent distribution morphism for transposed components
$\ga\<{\B,\A}\:T\B\#T\A\>T(\B\#\A)$ is naturally isomorphic to the
original morphism $\ga\<{\A,\B}\:T\A\#T\B\>T(\A\#\B)$. The
corresponding isomorphism (of morphisms) is provided by the pair
$\TPL{\si\<{T\A,T\B}\;,\;T\si\<{\A,\B}}$:
\[
\Dia(40,20){
\Put(0,0){T\B\#T\A}
\Put(40,0){T(\B\#\A)}
\Vec(10,0;1,0;20)\Put(20,3){\ga}
\Put(0,20){T\A\#T\B}
\Put(40,20){T(\A\#\B)}
\Vec(10,20;1,0;20)\Put(20,18){\ga}
\Vec(0,17;0,-1;13)\Put(4,10){\si T}
\Vec(40,17;0,-1;13)\Put(36,10){T\si}
}
\]

{\boldmath
$\al$-$\ga$ condition:}
Independent joint distribution for three components is ``naturally
invariant'' with respect to the order of parentheses. More
precisely, the morphisms
\[
 \ga\<{\A,\B\#\C} \. (i\<{T\A}\#\ga\<{\B,\C})
 \: T\A\#(T\B\#T\C) \> T(\A\#(\B\#\C))
\]
and
\[
\ga\<{\A\#\B,\C} \. (\ga\<{\A,\B}\#i\<{T\C})
 \: (T\A\#T\B)\#T\C \> T((\A\#\B)\#\C)
\]
(that take independent joint distributions for three components
with different order of parentheses) are naturally isomorphic via
$\TPL{\al\<{T\A,T\B,T\C}\;,\; T\al\<{\A,\B,\C}}$:
\[
\Dia(100,20){
\Put(0,0){T\A\#(T\B\#T\C)}
\Vec(17,0;1,0;18)\Put(25,3){T\#\ga}
\Put(50,0){T\A\#T(\B\#\C)}
\Vec(65,0;1,0;20)\Put(75,3){\ga}
\Put(100,0){T(\A\#(\B\#\C))}
\Put(0,20){(T\A\#T\B)\#T\C}
\Vec(17,20;1,0;18)\Put(25,18){\ga\#T}
\Put(50,20){T(\A\#\B)\#T\C}
\Vec(65,20;1,0;20)\Put(75,18){\ga}
\Put(100,20){T((\A\#\B)\#\C)}
\Vec(0,17;0,-1;13)\Put(4,10){\al T}
\Vec(100,17;0,-1;13)\Put(96,10){T\al}
}
\]

{\boldmath
$\mu$-$\ga$ condition.}
Independent joint distribution for results of mixing of two
second-order distributions may also be obtained by mixing the
corresponding second-order independent distributions:
\[
\Dia(80,20){
\Put(0,0){T\A\#T\B}
\Vec(10,0;1,0;60)\Put(40,3){\ga}
\Put(80,0){T(\A\#\B)}
\Put(0,20){TT\A\#TT\B}
\Vec(13,20;1,0;14)\Put(20,18){\ga T}
\Put(40,20){T(T\A\#T\B)}
\Vec(53,20;1,0;15)\Put(60,18){T\ga}
\Put(80,20){TT(\A\#\B)}
\Vec(0,17;0,-1;13)\Put(7,10){\mu\#\mu}
\Vec(80,17;0,-1;13)\Put(77,10){\mu}
}
\]

{\boldmath
$\et$-$\ga$ condition:}
Independent joint distribution for two ``singleton'' distributions
is just the corresponding ``singleton'' distribution on a product
space:
\[
\Dia(40,20){
\Put(0,0){T\A\#T\B}
\Vec(10,0;1,0;20)\Put(20,3){\ga}
\Put(40,0){T(\A\#\B)}
\Put(20,20){\A\#\B}
\Vec(17,17;-1,-1;14)\Put(5,14){\et\#\et}
\Vec(23,17;1,-1;14)\Put(31,14){\et}
}
\]

\section{ Informativeness of \\
information transformers}



\subsection{ Accuracy relation} 


In order to define informativeness relation we will need to
introduce first the following auxiliary notion.

\begin{definition}
We will say that $\HP$ is an \NW{accuracy} relation on an
IT-category $\CC$ if for any pair of objects $\A$ and $\B$ in
$\CC$ the set $\CC(\A,\B)$ of all ITs from $\A$ to $\B$ is
equipped with a partial order $\HP$ that satisfies the following
\NW{monotonicity} conditions:
\[
  a \HP a',\; b \HP b' \Imp a\.b \HP a'\.b',
\]
\[
  a \HP a',\; b \HP b' \Imp a*b \HP a'*b'.
\]
\end{definition}

Thus, the composition and the product are monotonous with respect
to the partial order $\HP$. For a pair of ITs $a,b\in\CC(\A,\B)$
we shall say that $a$ is more \NW{accurate} then $b$ whenever
$a\HP b$.

It obviously follows from the very definition of the operation
$\*$~\rf{3} and from the monotonicity conditions that the
operation $\*$ is monotone as well:
\[
  a \HP a',\; b \HP b' \Imp a\*b \HP a'\*b'.
\]

It is clear that for any IT-category there exists at least a
``trivial variant'' of the partial order $\HP$, namely, one can
choose an equality relation for $\HP$, that is, one can put $a\HP
b \iffdef a=b$. However, many categories of ITs (for example,
multivalued and fuzzy ITs) provide a ``natural'' choice of the
accuracy relation, which is different from the equality relation.

\subsection{ Definition of informativeness relation} 


Suppose $a\:\D\>\A$ and $b\:\D\>\B$ are two information transformers with a common source
$\D$. Assume that there exists an IT $c\:\A\>\B$ such that $c\.a=b$. Then any information that
can be obtained from $b$ can be obtained from $a$ as well (by attaching the IT $c$ next to $a$).
Thus, it is natural to consider the information transformer $a$ as being more informative than the IT
$b$ and also more informative than any IT less accurate than $b$.

Now we give the formal definition of the informativeness relation in the category of information
transformers.

\begin{definition}  
We shall say that an information transformer $a$ is more \NW{informative} (better) than $b$ if there
exists an information transformer $c$ such that $c\.a\HP b$, that is,
\[
  a \BI b \Iffdef \exists c \q c\.a\HP b.
\]
\end{definition}

It is easily verified that the informativeness relation $\BI$ is a preorder on the class of information
transformers in $\CC$. This preorder $\BI$ induces an equivalence relation $\EI$ in the following
way:
\[
  a \EI b \Iffdef a \BI b \And b \BI a.
\]

Obviously, the relation ``more informative'' extends the relation ``more accurate,'' that is,
\[
  a \HP b \imp a \BI b.
\]

\subsection{ Main properties of informativeness}


It can be easily verified that the informativeness relation $\BI$ satisfies the following natural
properties.

\begin{lem}1 
Consider all information transformers with a fixed source $\D$.

\begin{itemize}

\item[\It{a}] The identity information transformer $i\<\D$ is the most informative and the
terminal information transformer $z\<\D$ is the least informative:
\[
  \forall a \quad i\<\D \BI a \BI z\<\D.
\]

\item[\It{b}] Any information transformer $a\:\D\>\spc{B}\#\spc{C}$ is more informative
than its parts $\pi\<{\B,\C}\.a$ and $\nu\<{\B,\C}\.a$.
\item[\It{c}] The product $a*b$ is more informative than its components
\[
  a*b \BI a,b.
\]
\end{itemize}
\end{lem}

Furthermore, the informativeness relation is compatible with the composition and the product
operations.

\begin{lem}2 

\It{a} If $a \BI b$, then $a\.c \BI b\.c$.

\It{b} If $a \BI b$ and $c \BI e$, then $a*c \BI b*e$.
\end{lem}

\bprf\begin{prf}

\It{a} By assumption, there exists an information transformer $a'$ such that $a'\.a \HP b$.
In conjunction with monotonicity of composition it follows that
$a'\.(a\.c) \HP b\.c$.

\It{b} By assumption, there exist information transformers $a'$ and $c'$ such that $a'\.a\HP b$
and $c'\.c\HP e$.
Put $m=a'\*c'$. Then
\[
  m\.(a*c) = (a'\*c')\.(a*c) = (a'\.a)*(c'\.c) \HP b*e.\qed
\]
\noqed
\end{prf}\eprf

\subsection{ Structure of the family of \\
informativeness equivalence classes}


Let $a$ be some information transformer. We shall denote by $[a]$ the equivalence (with respect to
informativeness) class of $a$. We shall also use boldface for equivalence classes, that is, $a\in\EC
a$ is equivalent to $\EC a=[a]$.

\begin{trm}3 
Let $\mathfrak{J}(\D)$ be the family of informativeness equivalence classes for the class of all
information transformers with a fixed domain $\D$. The family $\mathfrak{J}(\D)$ forms a
partial ordered Abelian monoid $\TPL{\mathfrak{J}(\D),\BI,*,\EC 0}$ with the smallest
element $\EC 0$ and the largest element $\EC 1$, where
\[
  [a] \BI [b] \iffdef a \BI b,
\qquad
  [a] * [b] \eqdef [a*b],
\qquad
  \EC 0 \eqdef [z\<\D],
\qquad
  \EC 1 \eqdef [i\<\D].
\]

Moreover, the following properties hold:

\begin{itemize}
\item[\It{a}]  $\quad \EC {0*a=a}$,
\item[\It{b}]  $\quad \EC {1*a=1}$,
\item[\It{c}]  $\quad \EC {0 \LI a \LI 1}$,
\item[\It{d}]  $\quad \EC {a*b \BI a,b}$,
\item[\It{e}]  $\quad \EC {(a \BI b) \And (c \BI e) \Imp a*c \BI b*e}$.
\end{itemize}
\end{trm}

\bprf\begin{prf}
First note that the product operation $*$ is well-defined for equivalence classes, that is,
\[
  a \EI b \And c \EI e \Imp a*c \EI b*e.
\]
This immediately follows from Lemma{}~1.

Associativity of $*$ on $\mathfrak{J}(\D)$ follows immediately from the associativity of the
product of ITs (Definition{}~1, axiom{}~8%
).

Similarly, the commutativity of $*$ follows from the axiom{}~7
of Definition{}~1.

Now let us prove the properties.

The properties{}~\It{c} and{}~\It{d} immediately follow from the statements{}~\It{a} and
\It{c} of Proposition{}~1.

The Property{}~\It{e} is a consequence of Lemma{}~1.

To prove{}~\It{b} note that{}~\It{d} implies $\EC {1*a\BI1}$. Moreover, since $\EC 1$ is the
largest element, $\EC {1\BI1*a}$. Hence, $\EC {1*a=1}$.

Finally, let us prove{}~\It{a}. Assume that $a\in\EC a$. Note that in any category with terminal
object $\Z$, for every object $\R$ the product $\Z\#\R$ exists and is isomorphic to $\R$. Indeed,
it is easy to verify that $\TPL{\R,z\<\R,i\<\R}$ is a product (in terms of category theory) of
objects $\Z$ and $\R$.

Now let $j\:\R\>\Z\#\R$, namely, $j=z\<\R*i\<\R$. Since there exists a unique morphism
$b\:\D\>\Z\#\R$ such that $\pi\<\Z\.b=z\<\D$ and $\pi\<\R\.b=a$, we arrive at
$j\.a=b=z\<\D*a$. Thus, $\EC {a\BI0*a}$. The converse inequality $\EC {a\LI0*a}$ follows
from{}~\It{d}.
\end{prf}\eprf

\section{ Informativeness and synthesis of\\
 optimal
information transformers}


In this section, we consider an alternative (with respect to the above) approach to informativeness
comparison. This approach is based on treating information transformers as data sources for
decision-making problems.

\subsection{ Decision-making problems in categories of ITs}


Results of observations, obtained on real sources of information (e.g. indirect  measurements) are as
a rule unsuitable for straightforward interpretation. Typically it is assumed that observations suitable
for interpretation are those into a certain object $\U$ which in what follows will be called object of
\NW{interpretations} or object of \NW{decisions}.

By an \NW{interpretable} information transformer for signals from an object $\D$ we mean any
information transformer $a\:\D\>\U$.

It is usually thought that some interpretable information transformers are more suitable for
interpretation (of obtained results) than others. Namely, on a set $\CC(\D,\U)$ of information

transformers from $\D$ to $\U$, one defines some preorder relation $\BQ$, which specifies the
relative quality of various interpretable information transformers. Typically the relation $\BQ$ is
predetermined by the specific formulation of a problem of optimal information transformer synthesis
(that is, decision-making problem).

We shall say that an abstract \NW{decision-making problem} is determined by a triple
$\TPL{\D,\U,\BQ}$, where $\D$ is an object of studied (input) signals, $\U$ is an object of
decisions (or interpretations), and $\BQ$ is a preorder on the set $\CC(\D,\U)$.

We shall call a preorder $\BQ$ \NW{monotone} if for any $a,b\in\CC(\D,\U)$
\[
  a\HP b \imp a\BQ b,
\]
that is, more accurate IT provides better quality of interpretation.

For a given information transformer $a\:\D\>\A$ we shall also say that an IT $b$ \NW{reduces}
$a$ to an interpretable information transformer if $b\.a\:\D\>\U$, that is, if $b\:\A\>\U$. Such
an information transformer $b$ will be called a \NW{decision strategy}.

The set of all interpretable information transformers obtainable on the basis of $a\:\D\>\A$ will be
denoted $\UU_a\SS\CC(\D,\U)$:
\[
  \UU_a \eqdef \SET{b\.a|b\:\A\>\U}.
\]

We shall call a decision strategy $r\:\A\>\U$ \NW{optimal} (for the IT $a$ with respect to the
problem $\TPL{\D,\U,\BQ}$) if the IT $r\.a$ is a maximal element in $\UU_a$ with respect to
$\BQ$. Thus, a decision-making problem for a given information transformer $a$ is stated as the
problem of constructing optimal decision strategies.

\subsection{ Semantical informativeness}


The relation $\BQ$ induces a preorder relation $\BS$ on a class of information transformers
operating from $\D$ in the following way.

Assume that $a$ and $b$ are information transformers with the source $\D$, that is, $a\:\D\>\A$,
$b\:\D\>\B$. By definition, put
\[
  a \BS b \Iffdef
  \forall b'\:\B\>\U \q \exists a'\:\A\>\U \q
  a'\.a \BQ b'\.b.
\]
In other words, $a \BS b$ if for every interpretable information transformer $d$ derived from $b$
there exists an interpretable information transformer $c$ derived from $a$ such that $c \BQ d$, that
is,
\[
  a \BS b \Iff \forall d\in\UU_b \q \exists c\in\UU_a \q c \BQ d.
\]

It can easily be checked that the relation $\BS$ is a preorder relation.

It is natural to expect that if one information transformer is more informative than the other, then the
former will be better than the latter in any context. In other words, for any preorder $\BQ$ on the set
of interpretable information transformers the induced preorder $\BS$ is dominated by the
informativeness relation $\BI$ (that is, $\BS$ is weaker than $\BI$). The converse is also true.

\begin{definition} 

We shall say that an information transformer $a$ is \NW{semantically more informative} than $b$ if
for any interpretation object $\U$ and for any preorder $\BQ$ (on the set of interpretable
information transformers) $a\BS b$ for the induced preorder $\BS$.
\end{definition}

The following theorem is in some sense a ``completeness'' theorem, which establishes a relation
between ``structure'' ($b$ can be ``derived'' from $a$) and ``semantics'' ($a$ is uniformly better then
$b$ in decision-making problems).

\begin{trm}4 

For any information transformers $a$ and $b$ with a common source $\D$, information transformer
$a$ is more informative than $b$ if and only if $a$ is semantically more informative than $b$.
\end{trm}

\bprf\begin{prf}

Assume that $a\:\D\>\A$, $b\:\D\>\B$, and $a\BI b$, that is, there exists $c$ such that $c\.a\HP
b$. For any $b'\:\B\>\U$, take $a'=b'\.c\:\A\>\U$. Hence, by monotonicity of the preorder
$\BQ$, we have $a'\.a \BQ b'\.b$ and thus $a\BS b$.

Conversely, assume that for any decision object $\U$ and for any preorder $\BQ$ on $\CC(\D,\U)$
we have $a \BS b$. Then in particular $a\BS_b b$, where the semantical informativeness relation
$\BS_b$ corresponds to the decision problem with the decision object $\U_b=\B$ and the preorder
$\BQ_b$:
\[
  c\BQ_b d \Iffdef
  \( (d\HP b) \Rightarrow (c\HP b) \)
  .
\]

According to our hypothesis, for any IT $b'\:\B\>\B$ there exists an IT $a'\:\A\>\B$ such that
$a'\.a\BQ_b b'\.b$. By putting $b'=i\<{\B}$ we obtain $a'\.a\HP b$, that is, $a\BI b$.%
\end{prf}\eprf

Let us remark that the above proof relies heavily on the extreme extent of the class of decision
problems involved. This makes it possible to select for any given pair of ITs $a,b$ an appropriate
decision-making problem $\TPL{\D,\U_b,\BQ_b}$ in which the interpretation object $\U_b$ and
the preorder $\BQ_b$ depend on the IT $b$. However, in some cases it is possible to point out a
concrete (universal) decision-making problem such that
\[
  a\BI b \Iff a\BS b.
\]

\begin{trm}5 

Assume that for a given object $\D$ there exists an object $\widetilde\D$ such that for every
information transformer acting from $\D$ there exists an equivalent (with respect to informativeness)
IT acting from $\D$ to $\widetilde\D$, that is,
\[
  \forall\B \q \forall b\:\D\>\B \q \exists b'\:\D\>\widetilde\D \q b\EI b'.
\]
Let us choose the decision object $\U \eqdef \widetilde\D$ and the preorder $\BQ$, defined by
\[
  c\BQ d \Iffdef c\HP d.
\]
Then $a\BI b$ if and only if $a\BS b$.
\end{trm}

\bprf\begin{prf}
Suppose $a\:\D\>\A$, $b\:\D\>\B$, and $a \BS b$ for the preorder $\BQ$ defined in the
conditions of the theorem. Then
\[
  \forall b'\:\B\>\widetilde\D \q
  \exists a'\:\A\>\widetilde\D \q
  a'\.a \HP b'\.b .
\]

Let $\CE{b}\:\D\>\widetilde\D$ be an IT such that $\CE{b}\EI b$. Thus,
\[
  \exists d\:\B\>\widetilde\D \q
  d\.b \HP \CE{b}.
\]
Put $b'=d$. Then by assumption, there exists $a'$ such that $a'\.a\HP b'\.b\;(\HP\CE{b})$,
that is, $a\BI\CE{b}\;(\EI b)$ and therefore $a \BI b$.%
\end{prf}\eprf

Note that in general case an optimal decision strategy (if exists) can be nondeterministic. However,
in many cases it is sufficient to search optimal strategies among deterministic ITs. Indeed, in some
categories of information transformers the relation of ``accuracy'' satisfies the following condition:
every IT is dominated by some deterministic IT, that is, for every IT there exists a more accurate
deterministic IT.

\begin{prp}1 

Assume that $\TPL{\D,\U,\BQ}$ is a monotone decision-making problem in a category of ITs
$\CC$. Assume also that the following condition holds:
\[
  \forall c\in\Ar(\CC)\q \exists d\in\Ar(\DD)\q d \HP c.
\]
Then for any IT $a\:\D\>\R$ and for any decision strategy $r\:\R\>\U$ there exists a
deterministic strategy $r\<0\:\R\>\U$ such that $r\<0\.a \BQ r\.a$.
\end{prp}

\bprf\begin{prf}

According to the hypothesis, for the IT $r$ there exists a deterministic IT $r\<0 \HP r$. Hence, by
the monotonicity of composition, $r\<0\.a \HP r\.a$, and since the preorder $\BQ$ is monotone,
we obtain $r\<0\.a \BQ r\.a$.%
\end{prf}\eprf


\section{ Decision-making problems with \\
a prior information}


%

In this section we formulate in terms of categories of information transformers an analogy for the
classical problem of optimal decision strategy construction for decision \NW{problems with a prior
information} (or information \NW{a priori}). We also prove a counterpart of the \NW{Bayesian
principle} from the theory of statistical games [24, 36].
Like its statistical prototype it reduces the problem of constructing an optimal decision strategy to a
much simpler problem of finding an optimal decision for a \NW{posterior information} (or
information \NW{a posteriori}).

First we define in terms of categories of information transformers some necessary concepts, namely,
concepts of distribution, conditional information transformer, decision problem with a prior
information, and others.

\subsection{ Distributions in categories of ITs}

We shall say that a \NW{distribution} on an object $\A$ (in some fixed category of ITs $\CC$) is
any IT $f\:\Z\>\A$, where $\Z$ is the terminal object in $\CC$.

The concept of distribution corresponds to the general concept of an element of some object in a
category, namely, a morphism from the terminal object (see, e.g., [9]).

Any distribution of the form $h\:\Z\>\A\#\B$ will be called a \NW{joint distribution} on $\A$ and
$\B$. The projections $\pi\<{\A,\B}$ and $\nu\<{\A,\B}$ on the components $\A$ and $\B$
respectively, ``extract'' \NW{marginal distributions} $f$ and $g$ of the joint distribution $h$, that is,
\[
  f=\pi\<{\A,\B}\.h\:\Z\>\A,
\]
\[
  g=\nu\<{\A,\B}\.h\:\Z\>\B.
\]

We say that the components of a joint distribution $h\:\Z\>\A\#\B$ are \NW{independent}
whenever this joint distribution is completely determined by its marginal distributions, that is,
\[
  h=(\pi\<{\A,\B}\.h)*(\nu\<{\A,\B}\.h).
\]

Let $f$ be an arbitrary distribution on $\A$ and let $a\:\A\>\B$ be some information transformer.
Then the distribution $g=a\.f$ in some sense ``contains an information about $f$.'' This concept
can be expressed precisely of one consider the joint distribution \NW{generated} by the distribution
$f$ and the IT $a$:
\[
  h\:\Z\>\A\#\B, \qquad h=(i\<\A*a)\.f.
\]
Note, that the marginal distributions for $h$ coincide with $f$ and $g$, respectively. Indeed,
\[
  \pi\<{\A,\B}\.h = \pi\<{\A,\B}\.(i\<\A*a)\.f = i\<\A\.f = f,
\]
\[
  \nu\<{\A,\B}\.h = \nu\<{\A,\B}\.(i\<\A*a)\.f = a\.f = g.
\]

Let $h$ be a joint distribution on $\A\#\B$. We shall say that $a\:\A\>\B$ is a \NW{conditional}
IT for $h$ with respect to $\A$ whenever $h$ is generated by the marginal distribution
$\pi\<{\A,\B}\.h$ and the IT $a$, that is,
\[
  h = (i\<\A*a)\.\pi\<{\A,\B}\.h.
\]
Similarly, an IT $b\:\B\>\A$ such that
\[
  h = (b*i\<\B)\.\nu\<{\A,\B}\.h
\]
will be called a conditional IT for $h$ with respect to $\B$.

\subsection{ Bayesian decision-making problems}


Suppose that, like in Section~4, there are fixed two objects $\D$ and $\U$ in some category of ITs,
namely, the object of signals and the object of decisions, respectively. In a decision-making problem
with a prior distribution $f$ on $\D$ one fixes some preorder $\BQ_f$ on the set of joint
distributions on $\D\#\U$ for which $\D$-marginal distribution coincides with $f$.

Informally, any joint distribution $h$ on $\D\#\U$ of this kind can be considered as a joint
distribution of a studied signal (with the distribution $f=\pi\<{\D,\U}\.h$ on $\D$) and a decision
(with the distribution $g=\nu\<{\D,\U}\.h$ on $\U$). The preorder $\BQ_f$ determines how good
is the ``correlation'' between studied signals and decisions.

Formally, an abstract \NW{decision problem with a prior information} is determined by a quadruple
$\TPL{\D,\U,f,\BQ_f}$, where $\D$ is an object of studied signals, $\U$ is an object of decisions
(or interpretations), $f\:\Z\>\D$ is a \NW{prior} distribution (or distribution \NW{a priori}), and
$\BQ_f$ is a preorder on the set of ITs $h\:\Z\>\D\#\U$ that satisfy the condition
$\pi\<{\D,\U}\.h=f$.

Furthermore, suppose that there is a fixed IT $a\:\D\>\R$ (which determines a measurement; $\R$
can be called an object of \NW{observations}). An IT $r\:\R\>\U$ is called \NW{optimal} (for the
IT $a$ with respect to $\BQ_f$) if the distribution $(i*r\.a)\.f$ is a maximal element with
respect to $\BQ_f$. The set of all optimal information transformers is denoted $\Opt_f(a\.f)$.

\begin{trm}{6 {\rm(Bayesian principle)}} 

Let $f$ be a given prior distribution on $\D$, let $a\:\D\>\R$ be a fixed IT, and let $b\:\R\>\D$
be a conditional information transformer for $(i*a)\.f$ with respect to $\R$. Then the set of
optimal ITs $r\:\R\>\U$, namely, the set of optimal decision strategies for $f$ over $a\.f$
coincides with the set of optimal decision strategies for $b\.g$ over $g$, where $g=a\.f$:
\[
  \Opt_f(a\.f) = \Opt_{b\.g}(g).
\]
\end{trm}

\bprf\begin{prf}

Assume that $b\:\R\>\D$ is a conditional IT for $(i*a)\.f$ with respect to $\R$, that is,
\[
  (b*i)\.g=(i*a)\.f .
\]

Let $r\:\R\>\U$ be any IT. Then from the very definition of the conditional IT $b$ it follows
\[
  (i*r\.a)\.f=(b*r)\.g.
\]

Indeed,
\[
  (i*r\.a)\.f = (i\*r)\.(i*a)\.f = (i\*r)\.(b*i)\.g = (b*r)\.g.
\]

This distribution on $\D\#\U$ is maximal with respect to $\BQ_f$ if and only if $r$ is an optimal
decision strategy for $b\.g$ over $g$.%
\end{prf}\eprf

In a wide class of decision problems (e.g., in linear estimation problems) an optimal IT $r$ happens
to be deterministic and is specified by the ``deterministic part'' of the IT $b$.

For many categories of information transformers (for example, stochastic, multivalued, and fuzzy
ITs [13, 15, 24]) an optimal decision strategy $r$ can be constructed ``pointwise'' according to the following
scheme. For the given ``result of observation'' $y\in\R$ consider the conditional (posterior)
distribution $b(y)$ for $f$ under a fixed $g=y$, and put
\[
  r(y) \eqdef d_{b(y)},
\]
where $d_{b(y)}$ is an optimal decision with respect to the posterior distribution $b(y)$.

\section{ Examples of categories of \\
 information transformers}

In this section we present several examples of different classes
of information transformers. The major difference between them is
the way of representing uncertainty. In each case (except the
category stochastic linear ITs, which cannot be constructed as a
Kleisli category, but is a subcategory of one) we will mention the
corresponding: base category $\DD$, functor $T$, and natural
transformation $\ga$. ``Elementary'' definitions for these
categories may be found in
 [14, 21, 22].

\newcommand{\ST}{\cls{ST}}
\newcommand{\SLT}{\cls{SLT}}

\begingroup
\let\Par=\S
\renewcommand{\t}{\theta}
\renewcommand{\S}{{\mathfrak S}}
\renewcommand{\O}{\Omega}
\renewcommand{\o}{\omega}
\newcommand{\s}{\sigma}
\def\Int_#1{\int\limits_{#1}\!\!}
\newcommand{\MS}[1]{\TPL{\O_{#1},\S_{#1}}}

\subsection{ Stochastic ITs} 

Let $\DD = \Meas$, the category of measurable
spaces and measurable maps, $T\A$ is the space of all probability measures
on $\A,$ (details may be found in
 [37]) and $\ga\<{\A,\B}$ takes a pair of
distributions $\P, \Q$ to their product $\P \otimes \Q$, a distribution on
$\A\#\B$.

The category of stochastic information transformers $\ST$ consists of measurable spaces (as objects)
and transition probability functions (as morphisms, that is, information transformers)~[3, 4, 37].
Note that a classical statistical experiment (namely, a parametrized family of probability measures), a
statistics (namely, a measurable function of a sample of observations), and a decision strategy
(possibly, nondeterministic) can be represented by appropriate transition probability functions. Thus,
all the above concepts fit in well with this scheme.

Suppose $\A=\MS\A$ and $\B=\MS\B$ are two measurable spaces. A \NW{stochastic information
transformer} $a\:\A\>\B$ is determined by a real-valued function (transition probability
function{}~[3, 38, 39] 
$P_a(\o,B)$ of two arguments $\o\in\O_\A$, $B\in\S_\B$ that satisfy the following conditions:
\begin{itemize}
\item[(a)]
Given a fixed event $B\in\S_\B$, the map $P_a(\cdot,B)$ is a measurable function on $\O_\A$.
\item[(b)]
Given a fixed elementary event $\o\in\O_\A$, the map $P_a(\o,\cdot)$ is a probability measure
on $\MS\B$.
\end{itemize}

For a given stochastic information transformers $a\:\A\>\B$ and $b\:\B\>\C$ their
\NW{composition} $b\.a$ in the category $\ST$ corresponds to the transition probability function
(see [3,37]
\[
  P_{b\.a}(\o,C) \eqdef \Int_{\O_\B} P_b(\o',C)\, P_a(\o,d\o') \qquad
  \forall \o\in\O_\A, \quad \forall C\in\S_\C.
\]

The subcategory of \NW{deterministic} ITs is actually a category $\Meas$ of measurable spaces and
measurable maps. To every measurable map $\varphi\:\A\>\B$ there corresponds the transition
probability function
$$
  P_\varphi(\o,B) \eqdef  
\begin{cases}
 1,  &  \mbox{ if }  \varphi(\o)\in B , \cr
 0, & \mbox{ if } \varphi(\o)\not\in B; \qquad
  \forall\o\in\O_\A, \quad \forall B\in\S_\B.
\end{cases}
$$

The category $\Meas$ has products, namely, the product of measurable spaces $\A$ and $\B$ in
$\Meas$ is $\TPL{\A\#\B,\pi\<{\A,\B},\nu\<{\A,\B}}$, where
\[
  \A\#\B \eqdef \TPL{\O_\A\#\O_\B,\; \S_\A\otimes\S_\B},
\]
$\pi\<{\A,\B}$ and $\nu\<{\A,\B}$ are the projections from the Cartesian product $\O_\A\#\O_\B$
onto its components $\O_\A$ and $\O_\B$ respectively, and $\S_\A\otimes\S_\B$ is the product
of $\sigma$\_algebras $\S_\A$ and $\S_\B$.

For a given pair of ITs $a\:\D\>\A$ and $b\:\D\>\B$ with a common source we define their
\NW{product} $a*b\:\D\>\A\#\B$ so that for every $\o\in\O_\D$ the probability distribution
$P_{a*b}(\o,\cdot)$ on $\A\#\B$ is the product $\otimes$ of the distributions
$P_a(\o,\cdot)$ and $P_b(\o,\cdot)$, that is,
\[
  P_{a*b}(\o,\cdot) \eqdef P_a(\o,\cdot) \otimes P_b(\o,\cdot) \qquad
  \forall\o\in\O_\D.
\]
In other words (see, for example, [38]),
\ct{Neveu}),
the distribution $P_{a*b}$ is completely determined by the following condition:
\[
  P_{a*b}(\o,A\#B) \eqdef P_a(\o,A) \, P_b(\o,B) \qquad
  \forall\o\in\O_\D, \quad \forall A\in\S_\A, \quad \forall B\in\S_\B.
\]

The only obvious choice for the accuracy relation in the category of stochastic ITs seems to be the
equality relation.

Now let us demonstrate that the basic concepts of mathematical statistics are adequately described in
terms of this IT-category. Namely, we shall verify that the concepts of distribution, conditional
distribution, etc. (introduced above in terms of IT-categories), in the category of stochastic ITs lead
to the corresponding classical concepts.

Indeed, any probability distribution $Q$ on a given measurable space $\A=\MS\A$ is uniquely
determined by the morphism $f\:\Z\>\A$ from the terminal object
$\Z=\TPL{\{0\},\;\bigl\{\EMPTYSET,\{0\}\bigr\}}$ (a one-point measurable space) such
that
\[
  P_f(0,A)=Q(A) \qquad \forall A\in\S_\A.
\]
In what follows we shall omit the first argument in $P_f(0,A)$ and write just $P_f(A)$ instead.

A statistical experiment is described by a family of probability measures $Q_\t$ on some measurable
space $\B$. This family is usually parametrized by elements of a certain set $\O_\A$. Sometimes
(especially when statistical problems with a prior information are studied) it is additionally assumed
that the set $\O_\A$ is equipped by some $\s$-algebra $\S_\A$ and that $Q_\t(B)$ is a
measurable function of $\t\in\O_\A$ for all $B\in\S_\B$ (and thus, $Q_\t(B)$ is a transition
probability function [39]).
Therefore, such statistical experiment is determined by the stochastic information transformer
$a\:\A\>\B$, where
\[
  P_a(\t,B)=Q_\t(B)\qquad \forall \t\in\O_\A, \quad \forall B\in\S_\B.
\]

In the case when no $\s$-algebra on the set $\O_\A$ is specified, one can put
$\S_\A=\spc{P}(\O_\A)$, that is, the $\s$-algebra of all the subsets of the set $\O_\A$. It is clear
that in this case the function $P_a(\t,B)=Q_\t(B)$ is a measurable function of $\t\in\O_\A$ for
every fixed $B\in\S_\B$ and thus (being a transition probability function), is described by a
stochastic IT $a\:\A\>\B$.

Note also, that any statistic, being a measurable function, is represented by a certain deterministic IT.
Decision strategies also correspond to deterministic ITs. At the same time, nondeterministic (mixed)
decision strategies are adequately represented by stochastic information transformers of general kind.

Now, let $f$ be some fixed distribution on $\A$ and let $a\:\A\>\B$ be some IT. The joint
distribution  $h$ on $\A\#\B$, generated by $f$ and $a$ (from the IT-categorical point of view, see
Section{}~7) is
\[
  h = (i*a)\.f.
\]
It means that for every set $A\#B$, where $A\in\O_\A$ and $B\in\O_\B$,
\begin{eqnarray*}
  P_h(A\#B)
  &=&
  \Int_{\O_\A} P_{i*a}(\o,\,A\#B)\,P_f(d\o)
  \=
  \Int_{\O_\A} P_i(\o,A)\,P_a(\o,B)\,P_f(d\o)
  \=
  \Int_A P_a(\o,B)\,P_f(d\o).
\end{eqnarray*}
Thus we come to the well known classical expression for the generated joint distribution (see, for
example, [39]).

Now assume that $P_f$ is considered as some probability \NW{prior} distribution (or distribution
\NW{a priori}) on $\A$. Then for a given transition probability function $P_a$, a \NW{posterior}
(or \NW{conditional}) distribution $P_b(\o',\cdot)$ on $\A$ for a fixed $\o'\in\O_\B$ is
determined, accordingly to [39]
by a transition probability function $P_b(\o',A)$, $\o'\in\O_\B$, $A\in\S_\A$ such that
\[
  P_h(A\#B) = \Int_B P_b(\o',A)\,P_g(d\o')\qquad
  \forall A\in\S_\A, \quad \forall B\in\S_\B,
\]
where
\[
  P_g(B) = \Int_{\O_\A} P_a(\o,B)\,P_f(d\o) \qquad \forall B\in\S_\B.
\]
It is easily verified that in terms of ITs the above expressions have the following forms:
\[
  h=(b*i)\.g,
\]
where
\[
  g=a\.f.
\]
This shows, that the classical concept of conditional distribution is adequately described by the
concept of conditional IT in terms of categories of information transformers.
\endgroup

\begingroup
\renewcommand{\S}{\Sigma}

\subsection{ Linear stochastic ITs with additive noise}



As we will see this category of ITs cannot be constructed as a
Kleisli category, but is a subcategory the category of stochastic
ITs, examined above.

Suppose $\D$ and $\R$ are arbitrary finite-dimensional Euclidean spaces. We shall say that a linear
information transformer [11, 12]
(measurement model [40])
\ct{Pyt:ZadRed})
$a$ acting from $\D$ to $\R$
\[
  a\:\D\>\R,
\]
is determined by a pair
\[
  \TPL{A_a,\S_a}, \qquad A_a\:\D\>\R,\qquad \S_a\:\R\>\R, \quad \S_a\GE 0,
\]
where $A_a$ and $\S_a$ are linear maps.

Such pair $\TPL{A_a,\S_a}$ represents a statistical experiment of the form [40]
\[
  y = A_a x + \nu, \qquad x\in\D, \qquad y\in\R,
\]

where $\nu$ is a random vector in $\R$ with the zero mean and the correlation operator $\S_a$.

The \NW{composition} of two linear ITs $\TPL{A_a,\S_a}\:\D\>\A$ and
$\TPL{A_b,\S_b}\:\A\>\B$ is defined by
\[
  \TPL{A_b,\S_b}\.\TPL{A_a,\S_a} \eqdef \TPL{A_b A_a,\S_b+A_b\S_a A_b^*}.
\]

The composition corresponds to the consecutive connection of information transformers that have
independent random errors.

The \NW{product} of two information transformers
$$
 \TPL{A_a,\S_a}\:\D\>\A, \qquad \TPL{A_b,\S_b}\:\D\>\B
$$
is defined by:
\[
  \TPL{A_a,\S_a}*\TPL{A_b,\S_b} \eqdef
  \TPL{A_{a*b},\S_{a*b}} \: \D\>\A\#\B,
\]
where
\[
  A_{a*b}\:\D\>\A\#\B, \quad A_{a*b}x\eqdef\TPL{A_ax,A_bx},
\]
\[
  \S_{a*b}\:\A\#\B\>\A\#\B,\quad \S_{a*b}\TPL{x,y}\eqdef\TPL{\S_ax,\S_by}.
\]

This construction gives us the category $\SLT$ with the
subcategory of deterministic ITs is (isomorphic to) the category
of Euclidean spaces and linear maps. In this case a linear map
$A\:\D\>\R$ corresponds to the IT $\TPL{A,0}\:\D\>\R$.

As we have already mentioned this category of ITs cannot be
constructed as a Kleisli category over the category of finite
dimensional Euclidean spaces. Indeed we can not define a ``space
of distributions'' on some space $\A$ as a finite dimensional
linear space, and thus, can not define functor $T$ in the category
of finite dimensional Euclidean spaces. However, $\SLT$ may be
considered as a subcategory the category of stochastic ITs $\ST$,
examined above. Indeed, each Euclidean space may be considered as
a measurable space endowed with Borel $\sigma$-algebra. Finally,
we may consider an IT $a=\TPL{A_a,\S_a}\:\D\>\R$ (in $\SLT$) as
the transition probability, that takes an element $x\in\D$ to the
normal distribution $N(A_a x,\S_a)$ with the mean value $A_a x$
and the correlation operator $\S_a$. Routine verification shows,
that the composition and product operations are preserved under
such inclusion of $\SLT$ into $\ST$.

In addition to the trivial relation of accuracy (which coincides with the equality relation) one can
define the accuracy relation in the following way:
\[
  \TPL{A_a,\S_a}\HP\TPL{A_b,\S_b} \Iffdef A_a=A_b,\;\S_a\LE\S_b.
\]
However, it can be proved that the informativeness relations corresponding these different accuracy
preorders, actually coincide.

In the category of linear information transformers every equivalence class $[a]$ corresponds to a
pair $\TPL{\spc{Q},S}$,
where $\spc{Q}\SS\D$ is an Euclidean subspace and $S\:\spc{Q}\>\spc{Q}$ is nonnegative
definite operator, that is, $S\GE0$. In these terms
\[
  \TPL{\spc{Q}_1,S_1}\GE\TPL{\spc{Q}_2,S_2}
  \Iffdef
  \spc{Q}_1\supseteq\spc{Q}_2, \q S_1\upharpoonright\spc{Q}_2\LE S_2.
\]
Here $S_1\upharpoonright\spc{Q}_2$ (the restriction of $S_1$ on $\spc{Q}_2$) is defined by
the expression $S_1\upharpoonright\spc{Q}_2 \eqdef P_2I_1S_1P_1I_2$, where
$I_j\:\spc{Q}_j\>\D$ is the subspace inclusion, and $P_j\:\D\>\spc{Q}_j$ is the orthogonal
projection (cf. [11, 40]).

Note also that in the category of linear information transformers every IT is dominated (in the sense
of the preorder relation $\HP$) by a deterministic IT. Hence, according to Proposition{}~2, in any
monotone decision-making problem without loss of quality one can search optimal decision

strategies in the class of deterministic ITs.

It is shown in [12],
that in the category of linear ITs for any joint distribution there always exist conditional distributions.
Thus in problems with a prior information one can apply Bayesian principle. Its direct proof in the
category of linear ITs as well as the explicit expression for conditional information transformers can
be found in [12].
\endgroup

\begingroup

\subsection{ The category of sets as a category of ITs}


As a trivial example of IT-category we consider the category of
sets $\Set$, whose objects are sets and morphisms are maps. This
category has products, hence all the ITs are deterministic. In
fact this category is trivially a Kleisli category with identity
functor as functor $T$.

It is not hard to prove that for a given set $\D$, the class of equivalent informativeness for an IT $a$
with the set $\D$ being its domain, is completely determined by the following equivalence relation
$\EQ_a$ on $\D$:
\[
  x \EQ_a y \iffdef ax = ay \qquad \forall x,y\in\D.
\]
Furthermore, $a\BI b$ if and only if the equivalence relation $\EQ_a$ is \NW{finer} than $\EQ_b$,
that is,
\[
  a\BI b \Iff \forall x,y\in\D \;\; \(x \EQ_a y \imp x \EQ_b y\).
\]

Thus, the partially ordered monoid of equivalence classes for ITs with the source $\D$, is isomorphic
to the monoid of all equivalence relations on $\D$ equipped with the order ``finer'' and with the
product:
\[
  x\; (\EQ_a*\EQ_b)\; y \Iffdef \(x \EQ_a y,\; x \EQ_b y\) \qquad
  \forall x,y\in\D.
\]
\endgroup

\begingroup
\newcommand{\BIW}{\mathrel{\dot\BI}}     
\renewcommand{\P}{{\spc{P}}}

\subsection{ Multivalued ITs}


Let $\DD = \Set$, the category of sets, $T\A$ is the set of all
nonempty subsets of $\A$ and $\ga\<{\A,\B}$ takes pair of sets
$\P, \Q$ to their Cartesian product $\P\#\Q$, a subset of
$\A\#\B$. This leads us to the category $\MVT$ of multivalued ITs.
Detailed
study of this category may be found in [14].
Thus, the category $\MVT$ consists of sets as objects and of multivalued maps
(everywhere defined relations) as morphisms (information
transformers). Despite its simplicity, this class of ITs may be
convenient when stochastic description of measurement error is
inadequate.

So, a multivalued IT $a$ from $\D$ to $\R$
\[
  a\:\D\>\R
\]
is determined by a multivalued map, that is,
\[
  \forall x\in\D\quad ax\SS\R,\quad ax\neq\EMPTYSET.
\]

Define the \NW{composition} and the \NW{product} of multivalued ITs by the following
expressions:
\[
  (b\.a)(x) \eqdef \bigcup\SET{by | y\in ax},
\]

\[
  (a*b)(x) \eqdef ax \# bx.
\]

The subcategory of deterministic ITs is actually the category of sets $\Set$.

In addition to the trivial accuracy relation in the category of multivalued ITs one can put
\[
  a \HP b \Iffdef \forall x\in\D\; ax\SS bx.
\]

These two accuracy relations lead to different informativeness relations
[14],
called (strong) informativeness $\BI$ and weak informativeness $\BIW$.

For the both informativeness relations the classes of equivalent ITs with a fixed source $\D$ can be
described explicitly.

In the case of weak informativeness every class of equivalent ITs corresponds to a certain covering
$\P$ of the set $\D$, such that if $\P$ contains some set $B$ then it contains all its subsets:
\[
    \(\exists B\in\P\;(A\SS B)\) \Imp A\in\P.
\]

Moreover, a covering $\P_1$ is more (weakly) informative than $\P_2$ (namely, $\P_1$
corresponds to a class of more (weakly) informative ITs than $\P_2$) if $\P_1$ is contained in
$\P_2$, that is,
\[
  \P_1 \BIW \P_2 \Iffdef \P_1 \SS \P_2.
\]

In the case of (strong) informativeness every class of equivalent ITs corresponds to a covering $\P$
of the set $\D$, that satisfy the more complex condition:
\[
  \(\Big.
    \(\exists B\in\P\;A\SS B\)
    \;\&\;
    \(\exists \spc{B}\SS\P\;A=\cUp\spc{B}\)
  \)
  \Imp A\in\P.
\]

In this case
\begin{eqnarray*}
  \P_1 \BI \P_2
  &\Iffdef&
  \Big(
    \(\forall A\in\P_1 \q \exists B\in\P_2 \q A \SS B\)
  \\ &&
    \;\&\;
    \(\forall B\in\P_2 \q \exists \spc{A}\SS\P_1 \q B=\cUp\spc{A}\)
  \Big).
\end{eqnarray*}

In the category of multivalued information transformers every IT is dominated (in the sense of the
partial order $\HP$) by a deterministic IT. Thus, in the monotone decision-making problem one can
search optimal decision strategies in the class of deterministic ones.

For every joint distribution in the category of multivalued ITs there exist conditional
distributions [13].
Therefore, in decision problems with a prior information, the Bayesian approach can be effectively
applied.

\endgroup

\begingroup


\newcommand {\FMT}{\cls{FMT}}    

\newcommand {\FPT}{\cls{FPT}}    

\newcommand{\x}[1]{\mu_{{}_{\scriptstyle\vphantom\beta#1\!}}}

\renewcommand{\d}[1]{\delta_{#1}}             

\subsection{ Categories of fuzzy information transformers}



Here we define two categories of fuzzy information transformers
$\FMT$ and $\FPT$ that
correspond to different fuzzy theories [15]
Let $\DD = \Set$, $T\A$ is the set of all normalized
fuzzy subsets of $\A$ and $\ga\<{\A,\B}$ takes pair of fuzzy subsets $\P, \Q$ to
the fuzzy subset $\P\#\Q,$
($\P\#\Q$)($x,y) = \P(x) \otimes \Q(y), $where the operation
$\otimes$ may be defined in a variety of ways. The most common are the
minimum (the category $\FMT$) and product (the category $\FPT$) operations [15].

Objects of these categories are arbitrary sets and morphisms are
everywhere defined fuzzy maps, namely, maps that take an element
to a normed fuzzy set (a fuzzy set $A$ is normed if supremum of
its membership function $\x{A}$ is $1$). Thus, an information
transformer $a\:\A\>\B$ is defined by a membership function
$\x{ax}(y)$ which is interpreted as the grade of membership of an
element $y\in\B$ to a fuzzy set $ax$ for every element $x\in\A$.

The category $\FMT$.
Suppose $a\:\A\>\B$ and $b\:\B\>\C$ are some fuzzy maps. 
We define their \NW{composition}
$b\.a$ as follows: for every element $x\in\A$ put
\[
  \x{(b\.a)x}(z) \eqdef \sup_{y\in\B}\min\(\x{ax}(y),\;\x{by}(z)\).
\]

For a pairs of fuzzy information transformers $a\:\D\>\A$ and $b\:\D\>\B$ with the common
source $\D$, we define their \NW{product} as the IT that acts from $\D$ to the Cartesian product
$\A\#\B$, such that
\[
  \x{(a*b)x}(y,z) \eqdef \min\(\x{ax}(y),\;\x{by}(z)\).
\]

The category $\FPT$.
Define the \NW{composition} and the \NW{product} by the following expressions:
\[
  \x{(b\.a)x}(z) \eqdef \sup_{y\in\B}\(\x{ax}(y)\;\x{by}(z)\),
\]
\[
  \x{(a*b)x}(y,z) \eqdef \x{ax}(y)\;\x{by}(z).
\]

In the both defined above categories of fuzzy information transformers the subcategory of
\NW{deterministic} ITs is (isomorphic to) the category of sets $\Set$. Let $g\:\A\>\B$ be some
map (morphism in $\Set$). Define the corresponding fuzzy IT (namely, a fuzzy map, which is
obviously, everywhere defined) $\tilde{g}\:\A\>\B$ in the following way:
$$
  \x{\tilde{g}(x)}(y)\eqdef\d{g(x),y}
  =
  \begin{cases}
 1, & \mbox{ if } g(x)=y, \cr
 0, & \mbox{ if } g(x)\neq y.
\end{cases}
$$

Concerning the choice of accuracy relation, note, that in these
IT\_categories, like in the category of multivalued ITs, apart
from the trivial accuracy relation one can put for $a,b\:\A\>\B$
\[
  a \HP b \Iffdef \forall x\in\A\q \forall y\in\B\q \x{ax}(y)\LE \x{bx}(y).
\]

In each fuzzy IT-category these two choices lead to two different informativeness relations, namely
the strong and the weak ones.

Like in the categories of linear and multivalued ITs discussed above, monotone decision-making
problems admit restriction of the class of optimal decision strategies to deterministic ITs without loss
of quality.

It was shown in [15]
that for every joint distribution in the categories of fuzzy ITs there exist conditional distributions. It
allows Bayesian approach and makes use of Bayesian principle in decision problems with a prior
information for fuzzy ITs [15]
(see also [16--18]
where connections between fuzzy decision problems and the underlying fuzzy logic are studied).

In this section we introduced only several examples of IT-categories. Let us also remark that there is
an extensive literature that studies a wide spectrum of categories which are close in their structure to
IT-categories [35, 41--46].

\endgroup





\end{document}

\end{document}